\documentclass[journal]{IEEEtran}
\usepackage{hyperref}
\usepackage{caption, subcaption}
\usepackage{amsmath,amssymb,amsfonts}
\usepackage{color, colortbl}
\definecolor{Gray}{gray}{0.95}
\usepackage{stfloats}
\usepackage{tikz}
\usepackage{pgfplots}
\pgfplotsset{compat=1.11}
\usepackage{collcell}
\usepackage{hhline}
\usepackage{pgf}
\usepackage{multirow}
\def\colorModel{hsb} 
\newcommand\ColCell[1]{
  \pgfmathparse{#1<50?1:0}
    \ifnum\pgfmathresult=0\relax\color{black}\fi
  \pgfmathsetmacro\compA{0}      
  \pgfmathsetmacro\compB{#1/100} 
  \pgfmathsetmacro\compC{1}      
  \edef\x{\noexpand\centering\noexpand\cellcolor[\colorModel]{\compA,\compB,\compC}}\x #1
  } 
\newcolumntype{E}{>{\collectcell\ColCell}m{0.4cm}<{\endcollectcell}} 
\newcommand*\rot{\rotatebox{90}}
\begin{document}
\title{Multi Sensor-based Implicit User Identification}
\author{Muhammad Ahmad, Ali Kashif Bashir, Adil Mehmood Khan, Manuel Mazzara, Salvatore Distefano, Shahzad Sarfraz
\thanks{M. Ahmad and S. Sarfraz are with the Department of Computer Science, National University of Computer and Emerging Sciences, Islamabad, Chiniot-Faisalabad Campus, Chiniot 35400, Pakistan. e-mail: (mahmad00@gmail.com).}
\thanks{A. K. Bashir is with the Department of Computing and Mathematics, Manchester Metropolitan University, Manchester, United Kingdom.}
\thanks{A.M. Khan is with Machine Learning \& Knowledge Representation (MlKr) lab, Institute of Data Science \& AI, Innopolis University, Innopolis, 420500, Russia.}
\thanks{M. Mazzara is with the Institute of Software Development and Engineering, Innopolis University, Innopolis, 420500, Russia.}
\thanks{S. Distefano is with Dipartimento di Matematica e Informatica---MIFT, University of Messina, Messina, 98121, Italy.}
\thanks{Manuscript received September 2020.}}
\markboth{Journal of \LaTeX\ Class Files,~Vol.~, No.~, September~2020}%
{Shell \MakeLowercase{\textit{et al.}}: Bare Demo of IEEEtran.cls for IEEE Journals}
\maketitle
\begin{abstract}
Smartphones have ubiquitously integrated into our home and work environments, however, users normally rely on explicit but inefficient identification processes in a controlled environment. Therefore, when a device is stolen, a thief can have access to the owner's personal information and services against the stored passwords. As a result of this potential scenario, this work proposes an automatic legitimate user identification system based on gait biometrics extracted from user walking patterns captured by a smartphone. A set of preprocessing schemes is applied to calibrate noisy and invalid samples and augment the gait-induced time and frequency domain features, then further optimized using a non-linear unsupervised feature selection method. The selected features create an underlying gait biometric representation able to discriminate among individuals and identify them uniquely. Different classifiers (i.e. Support Vector Machine (SVM), K-Nearest Neighbors (KNN), Bagging, and Extreme Learning Machine (ELM)) are adopted to achieve accurate legitimate user identification. Extensive experiments on a group of $16$ individuals in an indoor environment show the effectiveness of the proposed solution: with $5$ to $70$ samples per window, KNN and bagging classifiers achieve $87-99\%$  accuracy, $82-98\%$ for ELM, and $81-94\%$ for SVM. The proposed pipeline achieves a $100\%$ true positive and $0\%$ false-negative rate for almost all classifiers.
\end{abstract}
\IEEEpeerreviewmaketitle
\section{Introduction}

\IEEEPARstart{R}{ecently}, the smartphone users exponentially increased to $3$ billion and are expected to further grow by several hundred million in coming years \footnote{Statista: www.statista.com}. Boosted by information and communication technologies (ICT), mobile, and personal devices are becoming a more and more powerful and thus trustworthy inseparable companion of our lives. Our cyber alter egos often store sensitive personal information such as photos, videos, bank account, credit, and debit card details, as well as cookies, passwords, and personal data managed by Internet applications. Such information should be kept confidential and not disclosed, preserving the smartphone from unauthorized accesses \cite{Ahmad2016}. Robust and reliable user identification methods can be an effective solution for achieving smartphone security \cite{Marsico2019}.

Considering that each individual has their own walking pattern, a gait (user walking) based identification mechanism has been proposed in \cite{Mahmad19, Ahmad19} using built-in sensors such as accelerometer and linear accelerometer. Gait-based legitimate user identification has more advantages than limits, including but not limited to unobtrusiveness, passive, implicit, concurrent, and continuous observability. However, the main advantage is the cost-effectiveness, exploiting built-in sensors without any additional hardware required for the identification, just walking with the smartphone. Furthermore, gait-based legitimate user identification avoids identification processing and delays during login by continuously operating in the background while the user is walking. It is also hard to violate since an attacker needs to exactly reproduce the smartphone owner's gait, which depends on their silhouette and activity, captured by several-different sensors. Gait-based legitimate user identification can also be used as one of the security levels in \textit{multilevel security systems} \cite{Damasevicius16} combining gait patterns with other security info in crime analysis. 

In recent years, several identification approaches have been proposed leveraging smartphone's sensors such as \cite{Yang2020EchoIAIA, Yang2020BubbleMapPM, Ayotte2020FastFA, LU2020101861}. For instance, \cite{Clarke2006, Shi2011} presented a method for continuous user identification implicitly. Explicit identification is performed only when there is important evidence of change in the user activity, which is not a real-life scenario in many cases. A method to directly compute the distance between pattern traces using the dynamic time warping algorithm is presented in \cite{DeLuca2012}. Sae, et.al., \cite{Sae2012} presents $22$ special touch patterns for user identification, most of which involve all five fingers simultaneously. The work \cite{Frank2013} studied the correlation between $22$ analytic features from touch traces and classified them using k-nearest neighbors and support vector machines.

Moreover, the idea behind the behavior-based model is that the person's habits are a set of attributes; therefore, each activity (event) correlates with two fundamental attributes: time and space. For instance, the works \cite{Rocha2011, Lima2011}, utilizes the user calls, schedules, GNSS, device battery level, user applications, and sensors for identification. The works \cite{Sabharwal2017, Mantyjarvi2005, Gafurov2009, Kale2002} proposed a multi-model-based continuous user identification. Whereas, the work \cite{Jakobsson2009} put forward another unique implicit user identification framework by using recorded phone call history and location for continuous user identification. 

\begin{table*}[!hbt]
    \caption{Summary of related work.}
    \scriptsize
    \resizebox{\textwidth}{!}{
    \centering
    \begin{tabular}{p{2.0cm}|p{5cm}|p{7cm}}
        \textbf{Reference} &  \textbf{Features} & \textbf{Methods}  \\ \hline
        Ahmad \cite{Ahmad2016} & Time Domain & SVM and KNN \\ \hline
        Ahmad \cite{Ahmad19} & Time and Frequency & Decision Tree, KNN, SVM  \\ \hline 
        Ahmad \cite{Ahmad2019} & Time and Frequency & Extreme Learning Machine  \\ \hline 
        Hughes \cite{Hughes16} & \multicolumn{2}{c}{Genetic Programming} \\ \hline 
        Derawi \cite{Derawi13} & Magnitude of the acceleration Weighted moving average filter, cycle detection, Manhattan distance & SVM  \\ \hline
        Davidson \cite{Davidson16} &  Gait, Time and Frequency & KNN and Random Forest \\ \hline
        Kobayashi \cite{Kobayashi11} & Cross-correlations of Fourier transform & Nearest means in Fisher discriminant space and majority voting \\ \hline  
        Thang \cite{Thang12} & Time and Frequency & Gait templates, DTW, SVM \\ \hline 
        Wolff \cite{Wolff13} & Variance in acceleration and orientation across $x, y, z$ & Gaussian distribution model \\ \hline
        Juefei-Xu \cite{Juefei12} & Accelerometer and gyroscope data & SVM and a time frequency spectrogram and a cyclo-stationary model \\ \hline 
        S. Sprager \cite{Sprager09} & Acceleration & SVM \\ \hline
        Pan \cite{Pan09} & Extrema in acceleration  & Difference-of-Gaussian and KNN \\ \hline 
        Kwapisz \cite{Kwapisz10} & Time domain & J48 and ANN \\ \hline
        Lin \cite{Lin14} & Spectral energy diagrams of pitch, roll,
        acceleration $x, y, z$ & $\alpha \beta$ filtering, Empirical Mode Decomposition, Fourier Transform, Linear Discriminant Analysis \\ \hline 
        Lu \cite{Lu14} & Time and Frequency & Gaussian Mixture
        Model and Universal Background Model \\ \hline 
        Johnston \cite{Johnston15} & Time and Frequency & MLP, Nivie Bayer, Random Forest\\ \hline
        Trivino  \cite{Trivino10} & Acceleration & FFSM and linguistic model \\ \hline
        Wang \cite{Liang03} & Domain specific & DTW distance\\ \hline 
        Rong  \cite{Rong07} & Acceleration & DTW\\ \hline
        Ailisto  \cite{Heikki05} & Averaged $x$ and $z$ signals & Correlation, Template Matching \\ \hline 
        Bachlin  \cite{Bachlin09} & FFT coefficients & FFT and ANOVA \\ \hline
        Hoang \cite{Hoang13} & Magnitude of the acceleration forces
        acting $x, y, z$ & Gait template matching\\ \hline
        Nickel \cite{Nickel11}& Mel and Bark frequency cepstral coefficients & SVM\\ \hline
    \end{tabular}}
    \label{Tab.1}
\end{table*}

The above-discussed works present several propositions, but to some extent, all these required additional information and source for user identification. Several works have been proposed to overcome these propositions, such as \cite{Casale2012} presented a gait-based user identification over biometric unobtrusive pattern. A geometric concept of a convex hull was utilized in $4$-layered architecture. One of the major limitations is non-user-friendliness, e.g., only works in specific and controlled environments. The works \cite{Mantyjarvi2005, Rong2007, Derawi2010, Gafurov2010, MDerawi2010, Bours2010} utilized portable devices based on gait signals acquired with a $3$-dimensional accelerometer, where the accelerometer was put on the user's belt only at the back. Whereby, \cite{Mantyjarvi2005} proposed a $3$-fold method based on data distribution statistics, correlation, and frequency domain features for user identification while the individuals are intentionally asked to walk with different speeds such as slow, normal, and fast. Mantyjarvi's work is novel but the major drawback is its limitations to not only walked by the same user but with very limited variations. 

Despite the success of the gait-based systems demonstrated by a relevant number of existing solutions, summarized above as well as in Table \ref{Tab.1}, there is still room for improving this approach, strongly depending on factors like physical changes i.e., aging, weight loss or gain, injury, shoes, clothes, carrying objects, orientation, and placement, walking surface, psychological states of an individual, stimulants, etc. All these factors significantly reduce the effectiveness of the gait-based system in real scenarios.

Considering the aforementioned scenarios, this work proposes a novel, non-intrusive, and automatic legitimate user identification system exploiting built-in smartphone motion dynamics captured by four different sensors namely, Accelerometer (AC), Linear Accelerometer (LAC), Gyroscope (GY), and Magnetometer (MM) sensors, able to overcome the limitations of existing solutions. To test the system, we first collect raw data from $16$ individuals walking with the smartphone freely placed in one of their pants pockets (\textit{front left, front right, back left, and back right}), then extracting relevant features from the raw data. To reduce the redundancy among such features a non-linear Extended Sammon Mapping Projection (ESMP) feature selection method is adopted, thus resulting in an underlying representation for the gait characteristics able to uniquely identify individuals. Finally, several machine learning classifiers e.g., SVM, KNN, Bagging, and ELM are implemented and compared to show the effectiveness-accuracy of the proposed gait-based legitimate user identification. In a nutshell, the following points are added in this research as compared to the previous works. 

\begin{enumerate}
\item Previous works only considered an activity/sub-activity-based user identification, however, our current research aims to propose a semi-controlled environment system in which we overcome the limitations of users' jeans style (loose or tight) and walking style (we intentionally asked users to walk differently in various times to investigate the ambulatory activity performed by each user). In this regard, our current work aims to investigate several research questions relevant to building a walking-based legitimate user identification system in real-life:
    \begin{itemize}
        \item{How to achieve real-time user identification in practice? Since our goal is to develop an algorithm that identifies the user in real-time, thus computation complexity is extremely important. System performance measurements ought to be considered to balance the trade-off between accuracy and computational cost.}
	    \item{Does the data variation affect the performance of the LUI process?}
	    \item{Does the Extended Sammon Mapping Projection (ESMP), a non-linear unsupervised feature selection method improve the identification accuracy more than the other existing and well-studied unsupervised feature selection methods such as Principal Component Analysis (PCA)? It is a known fact that the output of the smartphone sensor depends on the position of the smartphone while walking. This could result in a high within-class variance. Therefore, it is desirable to improve both the discriminatory power and achieve dimensionality reduction, by employing an optimum method. The advantages of the feature selection process are to avoid the curse of dimensionality, as well as to reduce the abundant, irrelevant, misleading, and noisy features, but above all, to be able to reduce the system's running cost pertaining to real-time applications. In addition to the above, effective feature selection can increase the accuracy of the resulting model.}
	    \item{Does kernel-based Extreme Learning Machine (KELM) an effective classifier for the non-linear signal-based user identification method then the state-of-the-art classification methods such as SVM, KNN, and Bagging? The reason to chose these classifiers, because these have been extensively utilized in the literature and rigorously analyzed for comparative analysis. Moreover, this work aims to show that the proposed pipeline can work well with a diverse set of classifiers.}
    \end{itemize}
    \item In our current work, we have invited $16$ users with $4$ activities i.e., user walked with the phone freely placed in one of their pants pockets i.e., front left, front right, back left, back right. 
    \item Previous works only considered a limited number of features extracted from two types of sensors, however, this work further involved the gait based features together with frequency and time domain features obtained through four different sensors which provide more confidence towards the ultimate results.
\end{enumerate}

The rest of the paper is structured as follows. Section \ref{sec:3} describes the proposed approach and main components of the gait-based legitimate user identification system. Section \ref{sec:4} reports on the experiments performed to demonstrate the effectiveness of the proposed approach by discussing the obtained results. Section \ref{sec:5} compares the proposed solution against state-of-the-art related works and finally, Section \ref{sec:6} concludes this paper with remarks and future research directions.

\section{Methodology}
\label{sec:3}

Smartphone sensors generate highly fluctuating time-series signals making legitimate user identification more challenging. Therefore, it is required to transform raw signals into relevant and meaningful features through a complex process including preprocessing, feature extraction, and selection. 

\subsection{Hardware and Preprocessing}
\label{sec:3.0}

Smartphones are equipped with a variety of sensors (hardware/software) that are useful for monitoring device movements. Some of them are AC, LAC, GY, and MM in which AC and LAC record the acceleration along three axes $(x, y, z)$ and can measure both the effects of Earth's gravity on the device and device movement, whereas, GY and MM eradicate the effects of Earth's gravity \cite{ Online}.

The smartphone (LG Nexus $4$ smartphone with Android Wear $v4.2$ OS in the experiments) runs a custom application gathering data from AC, LAC, GY, and MM sensors temporarily stored into a text file in a micro SD card and then transferred to a computer. The sensor sampling frequency is set to $50Hz$ and in total, $10$ minutes of raw samples were gathered from each individual ($8$ male and $8$ female) without any fixed protocol while carrying a smartphone in one of their pants pocket (i.e., \textit{front left, front right, back left, and back right}). It is worth mentioning that different smartphones have different sampling rates, therefore, in order to control the sensor reading process and for better generalization and validation, the sample rate is set to $50Hz$ instead to use the highest sampling rate within different smartphones \cite{Ahmad2019}.

For these reasons, we split the raw signals into windows ($5~ to ~105$ samples per window, respectively) to control the flow rate hence passing fewer data to the system. The selected sample size provides enough data to be able to extract quality features while ensuring a fast response. Before extracting relevant features, a third ordered moving average filter is applied in the preprocessing stage to reduce the sensor noise. 

\subsection{Feature Extraction}
\label{sec:3.1}

The raw AC, LAC, GY, and MM signals are processed to extract frequency, time, and gait features. These features are later combined into three feature vectors. Time-domain features solely consist of time and gait features vectors whereas the frequency and time feature vectors consist of both frequency and time domains.

The feature extraction process first analyzes sensors reading by applying time series modeling (i.e., Auto-regressive \cite{Cuomo97}, Moving average \cite{RUIZMEDINA2011292} and both auto-regressive and moving average models) to understand the behavior of users' physical patterns which reveals unusual observations and data patterns \cite{Ahmad2019}. Partial Auto-Correlation (PAC) and Auto-Correlation (AC) coefficients are used to identify the best model which revealed the pattern of each datum. Later each model is determined individually based on the characteristics of the theoretical PAC and AC. The best fit time-series model is calculated by estimating the parametric values based on the former model. Auto-regressive and moving average parameters are estimated through the box Jenkins model due to its flexibility for the inclusion of both models. The model and parameters need to be verified to ensure that the estimated results are statistically significant \cite{Mahmad19, Ahmad2019}. Our experiments revealed that the frequency and time domain features, including the coefficients from the time-series model, provide the best accuracies. Therefore, as listed in Table \ref{Tab.3}, gait, frequency, and time domain features are extracted from raw signals for each sensor individually. In total, $180$ features are extracted from each window.

\begin{table*}[!hbt]
    \centering
    \scriptsize
    \caption{Extracted features for gait-based legitimate user identification}
    \resizebox{\textwidth}{!}{
    \begin{tabular}{p{3cm}|p{12cm}} 
    \textbf{Feature Characteristics} & \textbf{Mathematical Reasoning} \\ \hline 
        Moving Variance &  $Var = \frac{1}{N(N-1)} \bigg(N \sum_{i=1}^N x_i^2 - \big(\sum_{i=1}^N x_i \big)^2 \bigg)$, where $x_i = a_z$ is accelerometer data along $z-axis$. Similarly computed along $x \& y-axis$ and for other sensors. \\ \hline 
        Moving Variance Intensity & $Var = \frac{1}{N(N-1)} \bigg(N \sum_{i=1}^N x_i^2 - \big(\sum_{i=1}^N x_i \big)^2 \bigg)$, where $x = \sqrt{a_x^2 + a_y^2 + a_z^2}$ is accelerometer data. Similarly computed for other sensors. \\ \hline 
        First eigenvalue of Moving Covariance  &  $E_1 = eig(Cov(a_x - g_x, a_y - g_y, a_z - g_z)$ where $a_x, a_y, a_z g_x, g_y, g_z$ are accelerometer and gyroscope readings. Similarly computed the other sensors. \\ \hline 
        Moving Covariance  & $E_a = eig_1 \bigg(Cov\big(a_x(1:N), a_y(1:N), a_z(1:N)\big)\bigg)$ where $a_x, a_y, a_z$ are accelerometer readings. Similarly computed the other sensors. \\ \hline 
        Moving energy &  $ME = \frac{1}{N} \sum_{i = 1}^N x_i^2$, where $x = a_z$, where $a_z$ is accelerometer reading along z-axis. Similarly computed for other sensors.\\ \hline 
        Moving Energy & $ME = \frac{1}{N} \sum_{i=1}^N (x_i -y_i)^2$, where $x = a_z, y = g_z$ are accelerometer and gyroscope data along z-axis. Similarly computed for other sensors. \\ \hline 
        Moving Energy of Orientation & $MEA = \frac{1}{N} \sum_{i+1}^N \phi_i^2$, where $\phi = \frac{arccos(a_x \times a_y)}{|a_x|\times |a_y|}$ is accelerometer readings. Similarly computed the other sensors. \\ \hline 
        Movement Intensity & $MI_a = \sqrt{a_x^2 + a_y^2 + a_x^2}$, where $a_x, a_y, a_z$ are accelerometer readings. Similarly computed the other sensors. \\ \hline
    \end{tabular}}
    \label{Tab.3}
\end{table*}

\subsection{Feature Selection}
\label{sec:3.2}

The sensor's output mainly depends upon the position of the smartphone, which may result in a high within-class variance \cite{Ahmad2019}. Therefore, it is required to enhance the discriminatory power of features that can achieve by deploying an optimum feature selection method. The feature selection process eliminates the irrelevant, abundant, noisy, and misleading features that reduce the system cost on run-time applications and improve the accuracy of the resulting model. A number of feature selection methods have been used for legitimate user identification. Filtering methods are interdependent to the classifier and depend on discriminating criteria i.e., maximum relevance and minimum redundancy. These methods are scalable, fast, and less computationally complex; however, ignore the interaction with the classifier. Wrapper methods utilize the classifier as a black box to obtain a subset based on their predictive power. The main drawback of the wrapper method is its dependency on the classifier which makes the classifier choice a key component. LDA and KLDA seek a linear combination of features. However, the number of dimensions depends on the number of classes which limits the use of such methods, especially for legitimate user identification.

To overcome the aforementioned issues, Extended Sammon Mapping Projection (ESMP) was first introduced in \cite{Ahmad2019} for smartphone-based physical activity recognition and legitimate user identification. ESMP is a nonlinear metric multi-dimensional scaling method that projects the high dimensional input space into lower dimensions while preserving the structure of inter-point distances. Let $d_{ij}$ and $d^*_{ij}$ be the Euclidean distance between two neighboring points $x_i$ and $x_j$ in input and mapped space, respectively. The Sammon stress error $E$ between the input and mapped space can be measured as explained in Equation \ref{eq.1} which is further optimized by the gradient descent method explained in Equation \ref{eq.2}.

\begin{equation}
    E = \frac{1}{\sum_{i = 1}^{n - 1} \sum_{j = i + 1}^{n} (d_{ij})} \times \sum_{i = 1}^{n-1} \sum_{j = i + 1}^{n} \frac{d^*_{ij} - d_{ij}}{d^*_{ij}}
    \label{eq.1}
\end{equation}

\begin{equation}
    x^*_{ik} (t + 1) =  x^*_{ik} (t) - \alpha \times \frac{\partial E(t)}{\partial x^*_{ik} (t)} 
    \label{eq.2}
\end{equation}
where $x^*_{ik}$ be the $k_{th}$ coordinate of $x_i$ in mapped space. The gradient descent methods have issues at inflation points in which the second-order derivatives appear to be quite small therefore the trade-off parameter $\alpha$ need to be set as a minimum (in our case $\alpha = [0.3-0.4]$ using grid search between $[0, 1]$. However, there is no guarantee that the given interval is to be optimal for all problems. Further details can be found in our previous work \cite{Ahmad2019}.

\subsection{Classifiers}
\label{sec:3.3}

The selected features are processed through Kernel Extreme Learning Machine (KELM), Bagging, SVM, and KNN. Several statistical measures are performed on the resulting false and true positive rates, ROC and accuracies for legitimate user identification are calculated for a different number of samples per window. 

\subsubsection{Kernel Extreme Learning Machine (KELM)}

ELM has fast learning speed and better generalization abilities than the other neural network frameworks. ELM randomly generates the input weights and bias with the help of a simple activation function. The tune-able activation functions were proposed to overcome the random assignments \cite{visapp19}. However, to find the suitable combinations for activation functions are still infancy. The KELM is used when the feature mapping functions of hidden neurons are unknown. However, the kernel parameters need to be tuned very carefully when it comes to real-time applications. Therefore, our current study explores the use of swarm optimizer to tune the kernel parameters. In this hierarchy, at the first hidden layer, all nonlinear piece-wise continuous functions can be used as hidden neurons as these parameters need not be tuned. Thus, for $N$ samples i.e., $(x_i, y_i) ~|~ x_i \in R^n$ and $y_i \in R^m$ where $ i \in 1, 2, 3, \dots , N$ and the output function can be represented as $f_L(x) = \sum_{i=1}^L \beta_i h_i(x) = h(x) \beta$, where $L$ is the number of hidden neurons, $\beta_i = [\beta_1, \beta_2, \beta_3, \dots, \beta_L]$ be the output weights among the output neurons and hidden layer. Finally, $h_i(x) = [h_1(x), h_2(x), h_3(x), \dots , h_L(x)]$ be the output vector that maps the input to the feature space. The least square solution that minimizes the error between training and output weights to boost the generalization capabilities can be represented in Eq. \ref{Eq.4}:

\begin{equation}
    \beta = H^T \bigg( \frac{1}{C} + H H^T \bigg)^{-1} \boldsymbol{T}
    \label{Eq.4}
\end{equation}
where $\boldsymbol{T}$, $H$ and $C$ are expected output, hidden layer output and regularization coefficients respectively. Thus, the training model output is expressed in Eq. \ref{Eq.5}:

\begin{equation}
    f(x) = h(x) H^T \bigg( \frac{1}{C} + H H^T \bigg)^{-1} \boldsymbol{T}
    \label{Eq.5}
\end{equation}

The output and kernel function $f(x)$ for unknown hidden mapping $h(x)$ can be written as $M = H H^T$ where $m_{i,j} = h(x_i)h(x_j) = \boldsymbol{k}(x_i, x_j)$. Thus, the final representation can be expressed as shown in Eq. \ref{Eq.6}:

\begin{equation}
    f(x) = [\boldsymbol{k}(x, x_1), \boldsymbol{k}(x, x_2), \dots, \boldsymbol{k}(x, x_N)] \bigg( \frac{1}{C} + M \bigg)^{-1} \boldsymbol{T}
    \label{Eq.6}
\end{equation}

\begin{equation}
    \boldsymbol{k}(x_i, x_j) = cos \bigg(\frac{||x - y||^2}{a} \bigg) \exp{\bigg(\frac{||x - y||^2}{b} \bigg)}
\end{equation}
where $\boldsymbol{k}(x_i, x_j)$ be the kernel function, $a, b$ are the tune-able parameters that plays an important role. 

\subsubsection{Support Vector Machine (SVM)}

Among the most popular methods for regression and classification problems, SVM is the most common classifier. SVM has been deployed for several real-world applications for instance bioinformatics, biometrics, cheminformatics, and remote sensing \cite{Ahmad2019A}. The SVM hierarchy works in two phases as the training examples are used to build the model for classification later the trained model is used to classify an unknown example. The tuning parameters of SVM are considered as key to success for any classification problems. These parameters include kernel and penalty parameters \cite{Ahmad2017}. The penalty is important to maintain a trade-off between maximizing the decision margin while minimizing the training error \cite{MAhmad2018}. Whereas, the kernel parameters are used to map the low dimensional input feature space to a higher dimensional feature space. These two parameters are labeled as a backbone to control the performance of SVM for any classification problem. SVM works while separating several known classes using the concept of hyperplanes and achieved remarkable results in linearly separable data examples.

\section{Results}
\label{sec:4}

The classifier's output indicates the ability to predict which user is walking while carrying the smartphone without considering its orientation, age, and gender. The $5$-fold cross-validation process is adopted to get meaningful and statistically significant results. The cross-validation process split the dataset into $5$ equal subsets in which $4$ subsets are selected to train the model and the remaining subset is selected to validate the model. This process is repeated $5$-times by picking a new subset (every time) as a validation set and remaining subsets are used as a training data lead to a total of $5$ experiments that weighted for the result.

\begin{figure}[!hbt]
    \centering
    \includegraphics[scale=0.45]{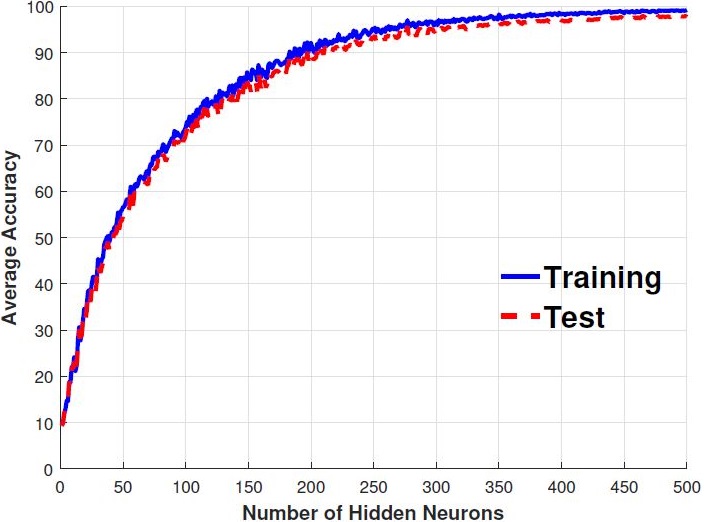}
    \caption{ELM: Accuracy vs Hidden Neurons}
    \label{Fig.101}
\end{figure}

The proposed pipeline has been evaluated against $3$ different types of classifiers such as KNN, SVM, and Bagging. The reason to chose these classifiers, because these have been extensively utilized in the literature and rigorously analyzed for comparative analysis. Moreover, this work aims to show that the proposed pipeline can work well with a diverse set of classifiers. For KELM, $[1-500]$ hidden neurons are selected, and SVM is evaluated with the polynomial kernel, similarly, for KNN, $k$ is set to $[2-20]$. For bagging, a tree-based method is used to train the classifier from a range of $[1-100]$ trees. All the parameters are adjusted carefully while setting up the experiments. The reason to provide the range is that because the number of samples in each round (samples per window) changes so the parameters need to be tuned in each round. For the main proposal, KELM classifier training and testing accuracy concerning the number of hidden neurons is presented in Figure \ref{Fig.101}.

For fare experimental evaluation of our proposed pipeline, several benchmark metrics are bring considered such as overall and average accuracies. For experimental validation and statistical analysis, true positive (TP), false positive (FP), true negative (TN), and false-negative (FN) is usually computed from the confusion matrix shown in Tables \ref{Tab.12}-\ref{Tab.15}.

\begin{table*}[!htb]
    \caption{Confusion Matrices for KELM Classifier.}
    \begin{subtable}{.5\linewidth}
      \centering
        \caption{\textbf{5 samples per window}}
        \newcommand\items{16}
        \arrayrulecolor{white}
        \resizebox{\columnwidth}{!}{
        \noindent\begin{tabular}{cc*{\items}{|E}|}
        \multicolumn{1}{c}{} &\multicolumn{1}{c}{} &\multicolumn{\items}{c}{Predicted}  \\ \hhline{~*\items{|-}|}  
        \multicolumn{1}{c}{} & 
        \multicolumn{1}{c}{} & 
        \multicolumn{1}{c}{\rot{1}} & 
        \multicolumn{1}{c}{\rot{2}} & 
        \multicolumn{1}{c}{\rot{3}} & 
        \multicolumn{1}{c}{\rot{4}} & 
        \multicolumn{1}{c}{\rot{5}} & 
        \multicolumn{1}{c}{\rot{6}} & 
        \multicolumn{1}{c}{\rot{7}} & 
        \multicolumn{1}{c}{\rot{8 }} & 
        \multicolumn{1}{c}{\rot{9}} & 
        \multicolumn{1}{c}{\rot{10}} & 
        \multicolumn{1}{c}{\rot{11}} & 
        \multicolumn{1}{c}{\rot{12}} & 
        \multicolumn{1}{c}{\rot{13}} & 
        \multicolumn{1}{c}{\rot{14}} & 
        \multicolumn{1}{c}{\rot{15}} & 
        \multicolumn{1}{c}{\rot{16 }} \\ \hhline{~*\items{|-}|}  
        \multirow{\items}{*}{\rotatebox{90}{Actual}} 
        &User 1& 53&	0&	0&	0&	0&	0&	0&	0&	0&	0&	0&	0&	0&	0&	0&	0 \\ \hhline{~*\items{|-}|}
        &User 2& 3&	47&	0&	4&	0&	1&	0&	1&	1&	1&	0&	0&	0&	2&	1&	1 \\ \hhline{~*\items{|-}|}
        &User 3& 0&	3&	50&	1&	0&	1&	0&	3&	1&	0&	0&	1&	0&	0&	2&	1 \\ \hhline{~*\items{|-}|}
        &User 4& 0&	0&	0&	49&	0&	0&	0&	0&	0&	2&	0&	0&	0&	1&	0&	1 \\ \hhline{~*\items{|-}|}
        &User 5& 1&	0&	0&	0&	50&	0&	0&	0&	2&	0&	0&	0&	0&	0&	0&	0 \\ \hhline{~*\items{|-}|}
        &User 6& 0&	0&	0&	0&	0&	58&	0&	1&	0&	0&	0&	0&	1&	0&	1&	1 \\ \hhline{~*\items{|-}|}
        &User 7& 0&	0&	1&	0&	0&	4&	52&	0&	1&	0&	0&	0&	1&	3&	1&	0 \\ \hhline{~*\items{|-}|}
        &User 8& 0&	0&	0&	0&	0&	2&	0&	46&	0&	2&	1&	0&	0&	0&	0&	2 \\ \hhline{~*\items{|-}|}
        &User 9& 2&	0&	0&	0&	1&	2&	0&	2&	35&	0&	0&	0&	0&	0&	3&	1 \\ \hhline{~*\items{|-}|}
        &User 10& 0&	2&	0&	0&	1&	3&	0&	1&	1&	55&	0&	0&	1&	2&	0&	0 \\ \hhline{~*\items{|-}|}
        &User 11& 0&	1&	0&	1&	0&	1&	1&	3&	0&	0&	40&	2&	0&	2&	0&	1 \\ \hhline{~*\items{|-}|}
        &User 12& 1&	1&	1&	1&	1&	0&	0&	7&	0&	0&	0&	30&	0&	0&	0&	2 \\ \hhline{~*\items{|-}|}
        &User 13& 1&	0&	0&	1&	0&	0&	3&	2&	4&	0&	0&	0&	32&	1&	0&	2 \\ \hhline{~*\items{|-}|}
        &User 14& 0&	0&	1&	0&	0&	1&	0&	0&	1&	0&	1&	0&	0&	59&	1&	2 \\ \hhline{~*\items{|-}|}
        &User 15& 0&	1&	2&	0&	1&	0&	1&	1&	0&	2&	0&	3&	0&	0&	40&	1 \\ \hhline{~*\items{|-}|}
        &User 16& 0&	1&	1&	0&	0&	0&	0&	0&	3&	1&	3&	0&	0&	1&	2&	32 \\ \hhline{~*\items{|-}|}
        \end{tabular}}
    \end{subtable}
    \begin{subtable}{.5\linewidth}
      \centering
        \caption{\textbf{105 samples per window}}
        \newcommand\items{16}   
        \arrayrulecolor{white} 
        \resizebox{\columnwidth}{!}{
        \noindent\begin{tabular}{cc*{\items}{|E}|}
        \multicolumn{1}{c}{} &\multicolumn{1}{c}{} &\multicolumn{\items}{c}{Predicted}  \\ \hhline{~*\items{|-}|}  
        \multicolumn{1}{c}{} & 
        \multicolumn{1}{c}{} & 
        \multicolumn{1}{c}{\rot{1}} & 
        \multicolumn{1}{c}{\rot{2}} & 
        \multicolumn{1}{c}{\rot{3}} & 
        \multicolumn{1}{c}{\rot{4}} & 
        \multicolumn{1}{c}{\rot{5}} & 
        \multicolumn{1}{c}{\rot{6}} & 
        \multicolumn{1}{c}{\rot{7}} & 
        \multicolumn{1}{c}{\rot{8 }} & 
        \multicolumn{1}{c}{\rot{9}} & 
        \multicolumn{1}{c}{\rot{10}} & 
        \multicolumn{1}{c}{\rot{11}} & 
        \multicolumn{1}{c}{\rot{12}} & 
        \multicolumn{1}{c}{\rot{13}} & 
        \multicolumn{1}{c}{\rot{14}} & 
        \multicolumn{1}{c}{\rot{15}} & 
        \multicolumn{1}{c}{\rot{16 }} \\ \hhline{~*\items{|-}|}  
        \multirow{\items}{*}{\rotatebox{90}{Actual}} 
        &User 1& 2111&	4&	4&	3&	3&	7&	2&	2&	0&	4&	6&	0&	3&	5&	0&	5 \\ \hhline{~*\items{|-}|}
        &User 2& 0&	2463&	4&	4&	1&	1&	0&	0&	0&	0&	11&	2&	0&	2&	8&	0 \\ \hhline{~*\items{|-}|}
        &User 3& 0&	0&	2538&	0&	5&	3&	1&	1&	4&	1&	2&	1&	2&	0&	1&	1 \\ \hhline{~*\items{|-}|}
        &User 4& 0&	11&	1&	2105&	3&	7&	2&	11&	0&	8&	6&	2&	16&	5&	1&	0 \\ \hhline{~*\items{|-}|}
        &User 5& 0&	0&	0&	0&	2154&	1&	0&	1&	0&	0&	0&	0&	1&	1&	0&	0 \\ \hhline{~*\items{|-}|}
        &User 6& 0&	7&	3&	0&	1&	2455&	3&	15&	1&	5&	0&	0&	1&	4&	0&	0 \\ \hhline{~*\items{|-}|}
        &User 7& 0&	0&	0&	1&	0&	1&	2550&	1&	0&	0&	0&	0&	0&	0&	0&	1 \\ \hhline{~*\items{|-}|}
        &User 8& 1&	2&	5&	2&	0&	2&	1&	2156&	0&	2&	0&	1&	4&	1&	1&	1 \\ \hhline{~*\items{|-}|}
        &User 9& 2&	5&	2&	1&	3&	1&	7&	3&	1831&	2&	3&	1&	3&	5&	6&	1 \\ \hhline{~*\items{|-}|}
        &User 10& 2&	0&	1&	1&	0&	3&	2&	4&	0&	2643&	14&	3&	1&	6&	4&	2 \\ \hhline{~*\items{|-}|}
        &User 11& 0&	1&	1&	4&	12&	3&	0&	2&	1&	4&	2093&	1&	0&	0&	1&	0 \\ \hhline{~*\items{|-}|}
        &User 12& 0&	0&	0&	0&	2&	0&	1&	0&	0&	1&	4&	1777&	0&	3&	1&	0 \\ \hhline{~*\items{|-}|}
        &User 13& 0&	3&	1&	4&	0&	2&	1&	0&	0&	1&	1&	0&	1861&	0&	0&	0 \\ \hhline{~*\items{|-}|}
        &User 14& 0&	4&	0&	3&	6&	3&	7&	4&	3&	3&	3&	1&	2&	2646&	0&	0 \\ \hhline{~*\items{|-}|}
        &User 15& 0&	6&	3&	2&	1&	0&	2&	1&	0&	5&	1&	1&	0&	3&	2092&	3 \\ \hhline{~*\items{|-}|}
        &User 16& 0&	1&	4&	0&	2&	6&	2&	2&	1&	4&	1&	0&	1&	8&	3&	1752 \\     \hhline{~*\items{|-}|}
        \end{tabular}}
    \end{subtable} 
    \label{Tab.12}
\end{table*}

\begin{table*}[!htb]
    \caption{Confusion Matrices for Bagging Classifier.}
    \begin{subtable}{.5\linewidth}
      \centering
        \caption{\textbf{5 samples per window}}
        \newcommand\items{16}   
        \arrayrulecolor{white} 
        \resizebox{\columnwidth}{!}{ \noindent\begin{tabular}{cc*{\items}{|E}|}
        \multicolumn{1}{c}{} &\multicolumn{1}{c}{} &\multicolumn{\items}{c}{Predicted}  \\ \hhline{~*\items{|-}|} 
        \multicolumn{1}{c}{} & 
        \multicolumn{1}{c}{} & 
        \multicolumn{1}{c}{\rot{1}} & 
        \multicolumn{1}{c}{\rot{2}} & 
        \multicolumn{1}{c}{\rot{3}} & 
        \multicolumn{1}{c}{\rot{4}} & 
        \multicolumn{1}{c}{\rot{5}} & 
        \multicolumn{1}{c}{\rot{6}} & 
        \multicolumn{1}{c}{\rot{7}} & 
        \multicolumn{1}{c}{\rot{8 }} & 
        \multicolumn{1}{c}{\rot{9}} & 
        \multicolumn{1}{c}{\rot{10}} & 
        \multicolumn{1}{c}{\rot{11}} & 
        \multicolumn{1}{c}{\rot{12}} & 
        \multicolumn{1}{c}{\rot{13}} & 
        \multicolumn{1}{c}{\rot{14}} & 
        \multicolumn{1}{c}{\rot{15}} & 
        \multicolumn{1}{c}{\rot{16 }} \\ \hhline{~*\items{|-}|}  
        \multirow{\items}{*}{\rotatebox{90}{Actual}} 
        &User 1& 50&	0&	0&	1&	0&	0&	0&	0&	0&	0&	0&	1&	0&	0&	0&	1\\ \hhline{~*\items{|-}|}
        &User 2& 0&	60&	1&	1&	0&	0&	0&	0&	0&	0&	0&	0&	0&	0&	0&	0\\ \hhline{~*\items{|-}|}
        &User 3& 0&	3&	59&	0&	0&	0&	0&	0&	0&	0&	0&	0&	0&	1&	0&	0\\ \hhline{~*\items{|-}|}
        &User 4& 0&	2&	0&	50&	0&	0&	0&	0&	0&	1&	0&	0&	0&	0&	0&	0\\ \hhline{~*\items{|-}|}
        &User 5& 0&	0&	0&	0&	53&	0&	0&	0&	0&	0&	0&	0&	0&	0&	0&	0\\ \hhline{~*\items{|-}|}
        &User 6& 1&	1&	0&	1&	0&	56&	1&	1&	1&	0&	0&	0&	0&	0&	0&	0\\ \hhline{~*\items{|-}|}
        &User 7& 0&	0&	0&	1&	0&	0&	61&	0&	1&	0&	0&	0&	0&	0&	0&	0\\ \hhline{~*\items{|-}|}
        &User 8& 0&	0&	2&	2&	1&	4&	1&	42&	0&	0&	0&	0&	0&	0&	0&	1 \\ \hhline{~*\items{|-}|}
        &User 9& 0&	0&	0&	0&	1&	3&	2&	0&	38&	0&	0&	0&	0&	1&	1&	0 \\ \hhline{~*\items{|-}|}
        &User 10& 0&	0&	0&	0&	0&	0&	0&	0&	0&	66&	0&	0&	0&	0&	0&	0 \\ \hhline{~*\items{|-}|}
        &User 11& 1&	0&	0&	0&	0&	0&	5&	1&	0&	1&	43&	0&	0&	1&	0&	0\\ \hhline{~*\items{|-}|}
        &User 12& 0&	0&	3&	1&	0&	0&	1&	0&	0&	0&	0&	38&	0&	1&	0&	0\\ \hhline{~*\items{|-}|}
        &User 13& 1&	0&	0&	1&	1&	0&	4&	2&	1&	3&	0&	1&	32&	0&	0&	0\\ \hhline{~*\items{|-}|}
        &User 14& 1&	2&	3&	0&	0&	0&	0&	1&	0&	1&	0&	0&	0&	58&	0&	0\\ \hhline{~*\items{|-}|}
        &User 15& 0&	3&	1&	0&	4&	0&	2&	1&	0&	2&	0&	0&	0&	0&	36&	3 \\ \hhline{~*\items{|-}|}
        &User 16& 0&	3&	0&	0&	0&	0&	0&	0&	0&	1&	2&	0&	3&	3&	0&	32\\ \hhline{~*\items{|-}|}
        \end{tabular}}
    \end{subtable}
    \begin{subtable}{.5\linewidth}
      \centering
        \caption{\textbf{105 samples per window}}
        \newcommand\items{16}   
        \arrayrulecolor{white} 
        \resizebox{\columnwidth}{!}{
        \noindent\begin{tabular}{cc*{\items}{|E}|}
        \multicolumn{1}{c}{} &\multicolumn{1}{c}{} &\multicolumn{\items}{c}{Predicted}  \\ \hhline{~*\items{|-}|}  
        \multicolumn{1}{c}{} & 
        \multicolumn{1}{c}{} & 
        \multicolumn{1}{c}{\rot{1}} & 
        \multicolumn{1}{c}{\rot{2}} & 
        \multicolumn{1}{c}{\rot{3}} & 
        \multicolumn{1}{c}{\rot{4}} & 
        \multicolumn{1}{c}{\rot{5}} & 
        \multicolumn{1}{c}{\rot{6}} & 
        \multicolumn{1}{c}{\rot{7}} & 
        \multicolumn{1}{c}{\rot{8 }} & 
        \multicolumn{1}{c}{\rot{9}} & 
        \multicolumn{1}{c}{\rot{10}} & 
        \multicolumn{1}{c}{\rot{11}} & 
        \multicolumn{1}{c}{\rot{12}} & 
        \multicolumn{1}{c}{\rot{13}} & 
        \multicolumn{1}{c}{\rot{14}} & 
        \multicolumn{1}{c}{\rot{15}} & 
        \multicolumn{1}{c}{\rot{16 }} \\ \hhline{~*\items{|-}|}  
        \multirow{\items}{*}{\rotatebox{90}{Actual}} 
        &User 1& 2158&	0&	0&	0&	0&	0&	0&	0&	0&	0&	0&	0&	0&	1&	0&	0\\ \hhline{~*\items{|-}|}
        &User 2& 0&	2495&	0&	0&	0&	0&	0&	0&	0&	0&	1&	0&	0&	0&	0&	0\\ \hhline{~*\items{|-}|}
        &User 3& 0&	0&	2558&	0&	0&	1&	0&	0&	0&	1&	0&	0&	0&	0&	0&	0\\ \hhline{~*\items{|-}|}
        &User 4& 3&	0&	0&	2173&	0&	0&	0&	1&	0&	1&	0&	0&	0&	0&	0&	0\\ \hhline{~*\items{|-}|}
        &User 5& 0&	0&	0&	0&	2158&	0&	0&	0&	0&	0&	0&	0&	0&	0&	0&	0\\ \hhline{~*\items{|-}|}
        &User 6& 0&	1&	0&	0&	0&	2488&	0&	2&	0&	1&	2&	0&	0&	1&	0&	0\\ \hhline{~*\items{|-}|}
        &User 7& 0&	0&	0&	0&	0&	0&	2554&	0&	0&	0&	0&	0&	0&	0&	0&	0\\ \hhline{~*\items{|-}|}
        &User 8& 0&	1&	0&	0&	0&	0&	0&	2177&	0&	0&	0&	0&	1&	0&	0&	0 \\ \hhline{~*\items{|-}|}
        &User 9& 2&	0&	0&	1&	0&	0&	2&	2&	1868&	1&	0&	0&	0&	0&	0&	0\\ \hhline{~*\items{|-}|}
        &User 10& 2&	0&	0&	0&	0&	0&	1&	0&	0&	2683&	0&	0&	0&	0&	0&	0 \\ \hhline{~*\items{|-}|}
        &User 11& 0&	0&	0&	0&	0&	0&	0&	0&	0&	1&	2122&	0&	0&	0&	0&	0\\ \hhline{~*\items{|-}|}
        &User 12& 0&	0&	0&	0&	0&	0&	0&	0&	0&	0&	0&	1789&	0&	0&	0&	0\\ \hhline{~*\items{|-}|}
        &User 13& 0&	0&	0&	0&	0&	0&	0&	0&	0&	0&	0&	0&	1874&	0&	0&	0\\ \hhline{~*\items{|-}|}
        &User 14& 0&	1&	0&	1&	0&	0&	0&	0&	0&	0&	0&	0&	0&	2683&	0&	0\\ \hhline{~*\items{|-}|}
        &User 15& 0&	1&	0&	0&	0&	1&	0&	0&	0&	0&	0&	0&	0&	0&	2118&	0 \\ \hhline{~*\items{|-}|}
        &User 16& 2&	0&	0&	1&	0&	0&	0&	1&	0&	0&	0&	0&	0&	1&	0&	1782\\ \hhline{~*\items{|-}|}
        \end{tabular}}
    \end{subtable} 
    \label{Tab.13}
\end{table*}

\begin{table*}[!htb]
    \caption{Confusion Matrices for KNN Classifier.}
    \begin{subtable}{.5\linewidth}
      \centering
        \caption{\textbf{5 samples per window}}
        \newcommand\items{16}   
        \arrayrulecolor{white} 
        \resizebox{\columnwidth}{!}{ \noindent\begin{tabular}{cc*{\items}{|E}|}
        \multicolumn{1}{c}{} &\multicolumn{1}{c}{} &\multicolumn{\items}{c}{Predicted}  \\ \hhline{~*\items{|-}|} 
        \multicolumn{1}{c}{} & 
        \multicolumn{1}{c}{} & 
        \multicolumn{1}{c}{\rot{1}} & 
        \multicolumn{1}{c}{\rot{2}} & 
        \multicolumn{1}{c}{\rot{3}} & 
        \multicolumn{1}{c}{\rot{4}} & 
        \multicolumn{1}{c}{\rot{5}} & 
        \multicolumn{1}{c}{\rot{6}} & 
        \multicolumn{1}{c}{\rot{7}} & 
        \multicolumn{1}{c}{\rot{8 }} & 
        \multicolumn{1}{c}{\rot{9}} & 
        \multicolumn{1}{c}{\rot{10}} & 
        \multicolumn{1}{c}{\rot{11}} & 
        \multicolumn{1}{c}{\rot{12}} & 
        \multicolumn{1}{c}{\rot{13}} & 
        \multicolumn{1}{c}{\rot{14}} & 
        \multicolumn{1}{c}{\rot{15}} & 
        \multicolumn{1}{c}{\rot{16 }} \\ \hhline{~*\items{|-}|}  
        \multirow{\items}{*}{\rotatebox{90}{Actual}} 
        &User 1& 52&	0&	0&	0&	0&	0&	0&	0&	0&	0&	0&	0&	0&	1&	0&	0\\ \hhline{~*\items{|-}|}
        &User 2& 0&	61&	0&	0&	0&	1&	0&	0&	0&	0&	0&	0&	0&	0&	0&	0\\ \hhline{~*\items{|-}|}
        &User 3& 0&	0&	62&	0&	0&	0&	0&	0&	0&	0&	0&	0&	0&	0&	1&	0\\ \hhline{~*\items{|-}|}
        &User 4& 0&	0&	0&	51&	0&	0&	1&	0&	0&	0&	0&	0&	0&	0&	1&	0\\ \hhline{~*\items{|-}|}
        &User 5& 0&	0&	0&	0&	52&	0&	0&	0&	0&	0&	0&	0&	0&	1&	0&	0\\ \hhline{~*\items{|-}|}
        &User 6& 0&	0&	0&	0&	0&	62&	0&	0&	0&	0&	0&	0&	0&	0&	0&	0\\ \hhline{~*\items{|-}|}
        &User 7& 0&	1&	0&	0&	0&	4&	58&	0&	0&	0&	0&	0&	0&	0&	0&	0\\ \hhline{~*\items{|-}|}
        &User 8& 0&	0&	0&	2&	0&	1&	5&	42&	0&	0&	0&	0&	0&	3&	0&	0 \\ \hhline{~*\items{|-}|}
        &User 9& 0&	0&	0&	1&	0&	2&	0&	0&	43&	0&	0&	0&	0&	0&	0&	0\\ \hhline{~*\items{|-}|}
        &User 10& 0&	0&	0&	0&	0&	0&	0&	0&	0&	66&	0&	0&	0&	0&	0&	0 \\ \hhline{~*\items{|-}|}
        &User 11& 0&	0&	1&	0&	0&	0&	2&	0&	0&	0&	49&	0&	0&	0&	0&	0\\ \hhline{~*\items{|-}|}
        &User 12& 1&	4&	2&	0&	0&	0&	0&	0&	0&	0&	1&	35&	1&	0&	0&	0\\ \hhline{~*\items{|-}|}
        &User 13& 0&	0&	0&	0&	0&	0&	2&	0&	0&	0&	0&	0&	42&	2&	0&	0\\ \hhline{~*\items{|-}|}
        &User 14& 1&	1&	0&	1&	0&	0&	1&	0&	0&	0&	0&	0&	0&	61&	1&	0\\ \hhline{~*\items{|-}|}
        &User 15& 0&	2&	0&	0&	0&	6&	1&	0&	0&	2&	1&	0&	0&	1&	39&	0 \\ \hhline{~*\items{|-}|}
        &User 16& 0&	0&	0&	1&	0&	2&	0&	0&	0&	1&	0&	1&	0&	0&	2&	37\\ \hhline{~*\items{|-}|}
        \end{tabular}}
    \end{subtable}
    \begin{subtable}{.5\linewidth}
      \centering
        \caption{\textbf{105 samples per window}}
        \newcommand\items{16}   
        \arrayrulecolor{white} 
        \resizebox{\columnwidth}{!}{
        \noindent\begin{tabular}{cc*{\items}{|E}|}
        \multicolumn{1}{c}{} &\multicolumn{1}{c}{} &\multicolumn{\items}{c}{Predicted}  \\ \hhline{~*\items{|-}|}  
        \multicolumn{1}{c}{} & 
        \multicolumn{1}{c}{} & 
        \multicolumn{1}{c}{\rot{1}} & 
        \multicolumn{1}{c}{\rot{2}} & 
        \multicolumn{1}{c}{\rot{3}} & 
        \multicolumn{1}{c}{\rot{4}} & 
        \multicolumn{1}{c}{\rot{5}} & 
        \multicolumn{1}{c}{\rot{6}} & 
        \multicolumn{1}{c}{\rot{7}} & 
        \multicolumn{1}{c}{\rot{8 }} & 
        \multicolumn{1}{c}{\rot{9}} & 
        \multicolumn{1}{c}{\rot{10}} & 
        \multicolumn{1}{c}{\rot{11}} & 
        \multicolumn{1}{c}{\rot{12}} & 
        \multicolumn{1}{c}{\rot{13}} & 
        \multicolumn{1}{c}{\rot{14}} & 
        \multicolumn{1}{c}{\rot{15}} & 
        \multicolumn{1}{c}{\rot{16 }} \\ \hhline{~*\items{|-}|}  
        \multirow{\items}{*}{\rotatebox{90}{Actual}} 
        &User 1& 2158&	0&	0&	0&	0&	0&	0&	0&	0&	0&	0&	0&	0&	1&	0&	0\\ 
        &User 1& 2153&	0&	0&	1&	0&	0&	0&	0&	0&	1&	1&	0&	1&	1&	1&	0\\ \hhline{~*\items{|-}|}
        &User 2& 0&	2495&	0&	0&	0&	0&	0&	0&	0&	0&	0&	1&	0&	0&	0&	0\\ \hhline{~*\items{|-}|}
        &User 3& 0&	1&	2558&	0&	0&	0&	0&	0&	0&	1&	0&	0&	0&	0&	0&	0\\ \hhline{~*\items{|-}|}
        &User 4& 0&	1&	0&	2173&	0&	1&	0&	0&	0&	2&	0&	1&	0&	0&	0&	0\\ \hhline{~*\items{|-}|}
        &User 5& 0&	1&	0&	0&	2157&	0&	0&	0&	0&	0&	0&	0&	0&	0&	0&	0\\ \hhline{~*\items{|-}|}
        &User 6& 0&	0&	0&	0&	0&	2494&	0&	0&	0&	1&	0&	0&	0&	0&	0&	0\\ \hhline{~*\items{|-}|}
        &User 7& 0&	0&	0&	0&	0&	0&	2554&	0&	0&	0&	0&	0&	0&	0&	0&	0\\ \hhline{~*\items{|-}|}
        &User 8& 0&	0&	0&	0&	0&	0&	0&	2179&	0&	0&	0&	0&	0&	0&	0&	0 \\ \hhline{~*\items{|-}|}
        &User 9& 0&	0&	0&	0&	0&	0&	0&	0&	1876&	0&	0&	0&	0&	0&	0&	0\\ \hhline{~*\items{|-}|}
        &User 10& 0&	0&	0&	0&	0&	1&	0&	0&	0&	2684&	1&	0&	0&	0&	0&	0 \\ \hhline{~*\items{|-}|}
        &User 11& 0&	1&	0&	0&	0&	0&	0&	0&	0&	0&	2122&	0&	0&	0&	0&	0\\ \hhline{~*\items{|-}|}
        &User 12& 0&	0&	0&	0&	0&	0&	0&	0&	0&	0&	0&	1789&	0&	0&	0&	0\\ \hhline{~*\items{|-}|}
        &User 13& 0&	0&	0&	0&	0&	0&	0&	0&	0&	0&	0&	0&	1874&	0&	0&	0\\ \hhline{~*\items{|-}|}
        &User 14& 0&	0&	0&	0&	0&	0&	0&	0&	1&	0&	0&	0&	0&	2684&	0&	0\\ \hhline{~*\items{|-}|}
        &User 15& 0&	0&	0&	0&	0&	0&	0&	1&	1&	0&	0&	0&	0&	0&	2118&	0 \\ \hhline{~*\items{|-}|}
        &User 16& 0&	0&	0&	0&	0&	0&	0&	1&	0&	0&	0&	0&	0&	0&	0&	1786\\ \hhline{~*\items{|-}|}
        \end{tabular}}
    \end{subtable} 
    \label{Tab.14}
\end{table*}

\begin{table*}[!htb]
    \caption{Confusion Matrices for SVM Classifier.}
    \begin{subtable}{.5\linewidth}
      \centering
        \caption{\textbf{5 samples per window}}
        \newcommand\items{16}   
        \arrayrulecolor{white} 
        \resizebox{\columnwidth}{!}{ \noindent\begin{tabular}{cc*{\items}{|E}|}
        \multicolumn{1}{c}{} &\multicolumn{1}{c}{} &\multicolumn{\items}{c}{Predicted}  \\ \hhline{~*\items{|-}|} 
        \multicolumn{1}{c}{} & 
        \multicolumn{1}{c}{} & 
        \multicolumn{1}{c}{\rot{1}} & 
        \multicolumn{1}{c}{\rot{2}} & 
        \multicolumn{1}{c}{\rot{3}} & 
        \multicolumn{1}{c}{\rot{4}} & 
        \multicolumn{1}{c}{\rot{5}} & 
        \multicolumn{1}{c}{\rot{6}} & 
        \multicolumn{1}{c}{\rot{7}} & 
        \multicolumn{1}{c}{\rot{8 }} & 
        \multicolumn{1}{c}{\rot{9}} & 
        \multicolumn{1}{c}{\rot{10}} & 
        \multicolumn{1}{c}{\rot{11}} & 
        \multicolumn{1}{c}{\rot{12}} & 
        \multicolumn{1}{c}{\rot{13}} & 
        \multicolumn{1}{c}{\rot{14}} & 
        \multicolumn{1}{c}{\rot{15}} & 
        \multicolumn{1}{c}{\rot{16 }} \\ \hhline{~*\items{|-}|}  
        \multirow{\items}{*}{\rotatebox{90}{Actual}} 
        &User 1&46&	0&	3&	0&	0&	0&	0&	0&	0&	0&	1&	0&	0&	2&	1&	0 \\ \hhline{~*\items{|-}|}  
        &User 2&0&	56&	4&	0&	0&	0&	0&	0&	0&	0&	0&	1&	0&	0&	0&	1 \\ \hhline{~*\items{|-}|} 
        &User 3&0&	0&	54&	0&	0&	5&	0&	0&	3&	1&	0&	0&	0&	0&	0&	0 \\ \hhline{~*\items{|-}|} 
        &User 4&0&	0&	0&	48&	0&	0&	0&	0&	0&	0&	0&	0&	0&	0&	4&	1 \\ \hhline{~*\items{|-}|} 
        &User 5&0&	0&	0&	0&	49&	0&	0&	1&	1&	0&	1&	0&	0&	0&	1&	0 \\ \hhline{~*\items{|-}|} 
        &User 6&0&	0&	0&	0&	0&	56&	5&	0&	0&	0&	1&	0&	0&	0&	0&	0 \\ \hhline{~*\items{|-}|}  
        &User 7&0&	0&	0&	0&	0&	1&	53&	1&	2&	0&	0&	0&	3&	1&	2&	0 \\ \hhline{~*\items{|-}|} 
        &User 8&0&	0&	0&	0&	0&	5&	1&	43&	0&	0&	0&	0&	0&	2&	1&	1 \\ \hhline{~*\items{|-}|}  
        &User 9&1&	0&	2&	0&	0&	3&	0&	0&	37&	0&	0&	0&	0&	3&	0&	0 \\ \hhline{~*\items{|-}|}  
        &User 10&0&	0&	0&	1&	0&	0&	0&	1&	0&	62&	0&	0&	2&	0&	0&	0 \\ \hhline{~*\items{|-}|}  
        &User 11&1&	1&	3&	0&	0&	1&	0&	1&	0&	0&	43&	0&	0&	1&	0&	1 \\ \hhline{~*\items{|-}|}  
        &User 12&0&	0&	6&	2&	0&	0&	0&	0&	1&	0&	1&	29&	0&	2&	3&	0 \\ \hhline{~*\items{|-}|} 
        &User 13&0&	0&	0&	1&	0&	1&	1&	0&	0&	2&	0&	0&	28&	11&	1&	1 \\ \hhline{~*\items{|-}|} 
        &User 14&0&	1&	0&	0&	0&	0&	0&	0&	1&	1&	0&	0&	0&	60&	3&	0 \\ \hhline{~*\items{|-}|}  
        &User 15&1&	3&	0&	1&	1&	2&	0&	0&	0&	0&	0&	0&	0&	0&	37&	7 \\ \hhline{~*\items{|-}|}  
        &User 16&0&	1&	1&	1&	0&	0&	1&	6&	0&	2&	0&	0&	4&	1&	1&	26 \\ \hhline{~*\items{|-}|} 
        \end{tabular}}
    \end{subtable}
    \begin{subtable}{.5\linewidth}
      \centering
        \caption{\textbf{105 samples per window}}
        \newcommand\items{16}   
        \arrayrulecolor{white} 
        \resizebox{\columnwidth}{!}{
        \noindent\begin{tabular}{cc*{\items}{|E}|}
        \multicolumn{1}{c}{} &\multicolumn{1}{c}{} &\multicolumn{\items}{c}{Predicted}  \\ \hhline{~*\items{|-}|}  
        \multicolumn{1}{c}{} & 
        \multicolumn{1}{c}{} & 
        \multicolumn{1}{c}{\rot{1}} & 
        \multicolumn{1}{c}{\rot{2}} & 
        \multicolumn{1}{c}{\rot{3}} & 
        \multicolumn{1}{c}{\rot{4}} & 
        \multicolumn{1}{c}{\rot{5}} & 
        \multicolumn{1}{c}{\rot{6}} & 
        \multicolumn{1}{c}{\rot{7}} & 
        \multicolumn{1}{c}{\rot{8 }} & 
        \multicolumn{1}{c}{\rot{9}} & 
        \multicolumn{1}{c}{\rot{10}} & 
        \multicolumn{1}{c}{\rot{11}} & 
        \multicolumn{1}{c}{\rot{12}} & 
        \multicolumn{1}{c}{\rot{13}} & 
        \multicolumn{1}{c}{\rot{14}} & 
        \multicolumn{1}{c}{\rot{15}} & 
        \multicolumn{1}{c}{\rot{16 }} \\ \hhline{~*\items{|-}|}  
        \multirow{\items}{*}{\rotatebox{90}{Actual}} 
        &User 1& 2056&	7&	4&	4&	0&	5&	4&	0&	1&	14&	3&	1&	7&	15&	3&	35 \\ \hhline{~*\items{|-}|}
        &User 2&0&	2420&	0&	12&	0&	4&	0&	20&	5&	2&	10&	3&	0&	14&	5&	1 \\ \hhline{~*\items{|-}|}
        &User 3&0&	6&	2489&	19&	6&	0&	1&	7&	2&	1&	4&	0&	2&	6&	3&	14 \\ \hhline{~*\items{|-}|}
        &User 4&0&	80&	7&	1690&	77&	32&	8&	50&	31&	70&	6&	1&	47&	7&	70&	2 \\ \hhline{~*\items{|-}|}
        &User 5&0&	32&	1&	0&	2102&	8&	0&	1&	0&	4&	6&	2&	0&	2&	0&	0 \\ \hhline{~*\items{|-}|}
        &User 6& 0&	5&	8&	1&	11&	2429&	4&	3&	0&	0&	13&	0&	1&	19&	0&	1 \\ \hhline{~*\items{|-}|}
        &User 7& 0&	0&	4&	0&	0&	1&	2537&	1&	1&	4&	4&	1&	0&	1&	0&	0 \\ \hhline{~*\items{|-}|}
        &User 8& 0&	2&	13&	13&	0&	13&	2&	2075&	3&	4&	1&	3&	2&	5&	43&	0 \\ \hhline{~*\items{|-}|}
        &User 9& 1&	48&	1&	23&	2	&0&	1&	16&	1651&	73&	1&	19&	8&	16&	13&	3 \\ \hhline{~*\items{|-}|}
        &User 10& 1&	12&	6&	10&	40&	10&	2&	72&	8&	2441&	48&	9&	3&	4&	9&	11 \\ \hhline{~*\items{|-}|}
        &User 11& 0&	7&	3&	4&	11&	3&	0&	0&	1&	3&	2053&	1&	8&	7&	0&	22 \\ \hhline{~*\items{|-}|}
        &User 12& 0&	0&	0&	2&	0&	6&	0&	2&	0&	2&	0&	1774&	0&	2&	1&	0 \\ \hhline{~*\items{|-}|}
        &User 13& 0&	0&	0&	0&	1&	0&	0&	9&	0&	0&	1&	1&	1857&	5&	0&	0 \\ \hhline{~*\items{|-}|}
        &User 14& 0&	15&	32&	14&	23&	9&	44&	5&	0&	4&	1&	13&	6&	2504&	12&	3 \\ \hhline{~*\items{|-}|}
        &User 15& 0&	1&	1&	0&	15&	7&	6&	9&	5&	62&	5&	4&	24&	6&	1975&	0 \\ \hhline{~*\items{|-}|}
        &User 16& 7&	3&	1&	0&	1&	9&	1&	6&	34&	4&	0&	0&	1&	28&	5&	1687 \\ \hhline{~*\items{|-}|}
        \end{tabular}}
    \end{subtable} 
    \label{Tab.15}
\end{table*}
Moreover, to validate the statistical significance, several statistical measures are considered such as Recall, Precision, and F1-score. Furthermore, this work carried out several statistical tests including but not limited to true positive rate (TPR), true negative rate (TNR), false-positive rate (FPR), and false-negative rate (FNR). Meanwhile, this study also used several other statistical measures to validate the performance of our proposed model as shown in Tables \ref{Tab.8}-\ref{Tab.9}. The FPR and TPR show two crucial aspects such as TPR and FPR show how usable and secure this would be as a legitimate user identification model. A low TPR shows that several legitimate attempts to identify would fail, thus making this too much of a burden to use, whereas a high FPR means illegitimate users could bypass the security and identification when they were not supposed to. Therefore, the ultimate goal of this work is to attain high TPR and low FNR as much as possible. The average statistical measures are computed as follows.

\arrayrulecolor{black}
\begin{table*}[!htb]
    \caption{Statistical Tests for KELM and Bagging Classifiers.}
    \begin{subtable}{.50\linewidth}
      \centering
        \caption{\textbf{KELM}}
        \resizebox{\columnwidth}{!}{\begin{tabular}{c|c|c|c|c|c|c}
        Users & Precision &	Sensitivity &	Specificity &	NPV &	FDR &	FOR \\ \hline
        User 1 & 1.0000 & 0.8689&	1.0000&	0.9903&	0.0000&	0.0097 \\ \hline 
        User 2 & 0.7581&	0.8393&	0.9818&	0.9890&	0.2419&	0.0110 \\ \hline 
        User 3 & 0.7937&	0.8929&	0.9842&	0.9926&	0.2063&	0.0074 \\ \hline 
        User 4 & 0.9245&	0.8596&	0.9951&	0.9903&	0.0755&	0.0097 \\ \hline 
        User 5 & 0.9434&	0.9259&	0.9964&	0.9952&	0.0566&	0.0048 \\ \hline 
        User 6 & 0.9355&	0.7945&	0.9950&	0.9816&	0.0645&	0.0184 \\ \hline 
        User 7 & 0.8254&	0.9123&	0.9866&	0.9939&	0.1746&	0.0061 \\ \hline 
        User 8 & 0.8679&	0.6866&	0.9914&	0.9745&	0.1321&	0.0255 \\ \hline 
        User 9 & 0.7609&	0.7143&	0.9867&	0.9832&	0.2391&	0.0168 \\ \hline 
        User 10 & 0.8333&	0.8730&	0.9865&	0.9901&	0.1667&	0.0099 \\ \hline 
        User 11 & 0.7692&	0.8889&	0.9856&	0.9939&	0.2308&	0.0061 \\ \hline 
        User 12 & 0.6818&	0.8333&	0.9834&	0.9928&	0.3182&	0.0072 \\ \hline 
        User 13 & 0.6957&	0.9143&	0.9834&	0.9964&	0.3043&	0.0036 \\ \hline 
        User 14 & 0.8939&	0.8310&	0.9913&	0.9852&	0.1061&	0.0148 \\ \hline 
        User 15 & 0.7692&	0.7843&	0.9855&	0.9867&	0.2308&	0.0133 \\ \hline 
        User 16 & 0.7273&	0.6809&	0.9856&	0.9820&	0.2727&	0.0180 \\ \hline 
    \end{tabular}}
    \end{subtable} \ \ \ \ \
    \begin{subtable}{.5\linewidth}
      \centering
        \caption{\textbf{Bagging}}
        \resizebox{\columnwidth}{!}{\begin{tabular}{c|c|c|c|c|c|c}
        Users & Precision &	Sensitivity &	Specificity &	NPV &	FDR &	FOR \\ \hline
        User 1 & 0.9434&	0.9259&	0.9964&	0.9952&	0.0566&	0.0048 \\ \hline
        User 2 & 0.9677&	0.8108&	0.9975&	0.9828&	0.0323&	0.0172 \\ \hline
        User 3 & 0.9365&	0.8551&	0.9951&	0.9877&	0.0635&	0.0123 \\ \hline
        User 4 & 0.9434&	0.8621&	0.9963&	0.9903&	0.0566&	0.0097 \\ \hline
        User 5 & 1.0000&	0.8833&	1.0000&	0.9915&	0.0000&	0.0085 \\ \hline
        User 6 & 0.9032&	0.8889&	0.9926&	0.9914&	0.0968&	0.0086 \\ \hline
        User 7 & 0.9683&	0.7922&	0.9975&	0.9804&	0.0317&	0.0196 \\ \hline
        User 8 & 0.7925&	0.8750&	0.9867&	0.9927&	0.2075&	0.0073 \\ \hline
        User 9 & 0.8261&	0.9268&	0.9904&	0.9964&	0.1739&	0.0036 \\ \hline
        User 10 & 1.0000&	0.8800&	1.0000&	0.9889&	0.0000&	0.0111 \\ \hline
        User 11 & 0.8269&	0.9556&	0.9892&	0.9976&	0.1731&	0.0024 \\ \hline
        User 12 & 0.8636&	0.9500&	0.9928&	0.9976&	0.1364&	0.0024 \\ \hline
        User 13 & 0.6957&	0.9143&	0.9834&	0.9964&	0.3043&	0.0036 \\ \hline
        User 14 & 0.8788&	0.8923&	0.9902&	0.9914&	0.1212&	0.0086 \\ \hline
        User 15 & 0.6923&	0.9730&	0.9810&	0.9988&	0.3077&	0.0012 \\ \hline
        User 16 & 0.7273&	0.8649&	0.9857&	0.9940&	0.2727&	0.0060 \\ \hline
    \end{tabular}}
    \end{subtable} 
    \label{Tab.8}
\end{table*}

\begin{table*}[!htb]
    \caption{Statistical Tests for KNN and SVM Classifiers.}
    \begin{subtable}{.5\linewidth}
      \centering
        \caption{\textbf{KNN}}
        \resizebox{\columnwidth}{!}{\begin{tabular}{c|c|c|c|c|c|c}
        Users & Precision &	Sensitivity &	Specificity &	NPV &	FDR &	FOR \\ \hline
        User 1 & 0.9811&	0.9630&	0.9988&	0.9976&	0.0189&	0.0024 \\ \hline
        User 2 & 0.9839&	0.8841&	0.9988&	0.9902&	0.0161&	0.0098 \\ \hline
        User 3 & 0.9841&	0.9538&	0.9988&	0.9963&	0.0159&	0.0037 \\ \hline
        User 4 & 0.9623&	0.9107&	0.9976&	0.9939&	0.0377&	0.0061 \\ \hline
        User 5 & 0.9811&	1.0000&	0.9988&	1.0000&	0.0189&	0.0000 \\ \hline
        User 6 & 1.0000&	0.7949&	1.0000&	0.9804&	0.0000&	0.0196 \\ \hline
        User 7 & 0.9206&	0.8286&	0.9938&	0.9853&	0.0794&	0.0147 \\ \hline
        User 8 & 0.7925&	1.0000&	0.9868&	1.0000&	0.2075&	0.0000 \\ \hline
        User 9 & 0.9348&	1.0000&	0.9964&	1.0000&	0.0652&	0.0000 \\ \hline
        User 10 & 1.0000&	0.9565&	1.0000&	0.9963&	0.0000&	0.0037 \\ \hline
        User 11 & 0.9423&	0.9608&	0.9964&	0.9976&	0.0577&	0.0024 \\ \hline
        User 12 & 0.7955&	0.9722&	0.9893&	0.9988&	0.2045&	0.0012 \\ \hline
        User 13 & 0.9130&	0.9767&	0.9952&	0.9988&	0.0870&	0.0012 \\ \hline
        User 14 & 0.9242&	0.8841&	0.9938&	0.9901&	0.0758&	0.0099 \\ \hline
        User 15 & 0.7500&	0.8864&	0.9844&	0.9939&	0.2500&	0.0061 \\ \hline
        User 16 & 0.8409&	1.0000&	0.9917&	1.0000&	0.1591&	0.0000 \\ \hline
    \end{tabular}}
    \end{subtable} \ \ \ \ \
    \begin{subtable}{.5\linewidth}
      \centering
        \caption{\textbf{SVM}}
        \resizebox{\columnwidth}{!}{\begin{tabular}{c|c|c|c|c|c|c}
        Users & Precision &	Sensitivity &	Specificity &	NPV &	FDR &	FOR \\ \hline
        User 1 &  0.8679 &	0.9388 &	0.9916 &	0.9964 &	0.1321 &	0.0036 \\ \hline 
        User 2 & 0.9032 &	0.9032 &	0.9926 &	0.9926 &	0.0968 &	0.0074 \\ \hline 
        User 3 & 	0.8571 &	0.7397 &	0.9888 &	0.9767 &	0.1429 &	0.0233\\ \hline 
        User 4 &  0.9057 &	0.8889 &	0.9939 &	0.9927 &	0.0943 &	0.0073\\ \hline 
        User 5 &  0.9245 &	0.9800 &	0.9952 &	0.9988 &	0.0755 &	0.0012 \\ \hline 
        User 6 &  0.9032 &	0.7568 &	0.9925 &	0.9779 &	0.0968 &	0.0221\\ \hline 
        User 7 &  0.8413 &	0.8689 &	0.9878 &	0.9902 &	0.1587 &	0.0098\\ \hline 
        User 8 &  0.8113 &	0.8113 &	0.9879 &	0.9879 &	0.1887 &	0.0121\\ \hline 
        User 9 &  0.8043 &	0.8222 &	0.9892 &	0.9904 &	0.1957 &	0.0096\\ \hline 
        User 10 &  0.9394 &	0.9118 &	0.9951 &	0.9926 &	0.0606 &	0.0074\\ \hline 
        User 11 &  0.8269 &	0.9149 &	0.9892 &	0.9952 &	0.1731 &	0.0048\\ \hline 
        User 12 &  0.6591 &	0.9667 &	0.9823 &	0.9988 &	0.3409 &	0.0012\\ \hline 
        User 13 &  0.6087 &	0.7568 &	0.9786 &	0.9892 &	0.3913 &	0.0108\\ \hline 
        User 14 &  0.9091 &	0.7229 &	0.9925 &	0.9717 &	0.0909 &	0.0283\\ \hline 
        User 15 &  0.7115 &	0.6852 &	0.9818 &	0.9794 &	0.2885 &	0.0206\\ \hline 
        User 16 &  0.5909 &	0.6842 &	0.9786 &	0.9856 &	0.4091 &	0.0144 \\ \hline 
    \end{tabular}}
    \end{subtable} 
    \label{Tab.9}
\end{table*}

\begin{equation}
    Average ~Accuracy = \frac{1}{M}\sum_{i = 1}^M \frac{TP_i + TN_i}{TP_i + TN_i + FP_i + TN_i}
\end{equation}

\begin{equation}
    Overall ~Accuracy = \frac{1}{M} \sum_{i = 1}^M TP_i
\end{equation}
where $M$ be the number of users.

\begin{equation}
    Precision = \frac{1}{M} \sum_{i = 1}^M \frac{TP_i}{TP_i + FP_i}
\end{equation}

\begin{equation}
    Sensitivity = \frac{1}{M} \sum_{i = 1}^M \frac{TP_i}{TP_i + FN_i}
\end{equation}

\begin{equation}
    Specificity = \frac{1}{M} \sum_{i = 1}^M \frac{TN_i}{TN_i + FP_i}
\end{equation}

\begin{equation}
    NPV = \frac{1}{M} \sum_{i = 1}^M \frac{TN_i}{TN_i + FN_i}
\end{equation}

\begin{equation}
    FDR = \frac{1}{M} \sum_{i = 1}^M \frac{FP_i}{FP_i + TP_i}
\end{equation}

\begin{equation}
    FOR = \frac{1}{M} \sum_{i = 1}^M \frac{FN_i}{FN_i + TN_i}
\end{equation}
where NPV stands for Negative Predictive Values, FDR is False Discovery Rate, and FOR is False Omission Rate. NPV is the score of the negative statistical results based on TN and FN values across the users used in this study. FDR and FOR are usually used in multiple hypothesis testing to make sure the multiple comparisons. FOR can be computed by taking the complement of NPV values or another way around, it can be measured using TN and FN values. FOR is used to measure the rate of false negatives, which are incorrectly rejected whereas, FDR measures the actual positives which were incorrectly identified.

\begin{figure}[!hbt]
    \centering
    \includegraphics[scale=0.35]{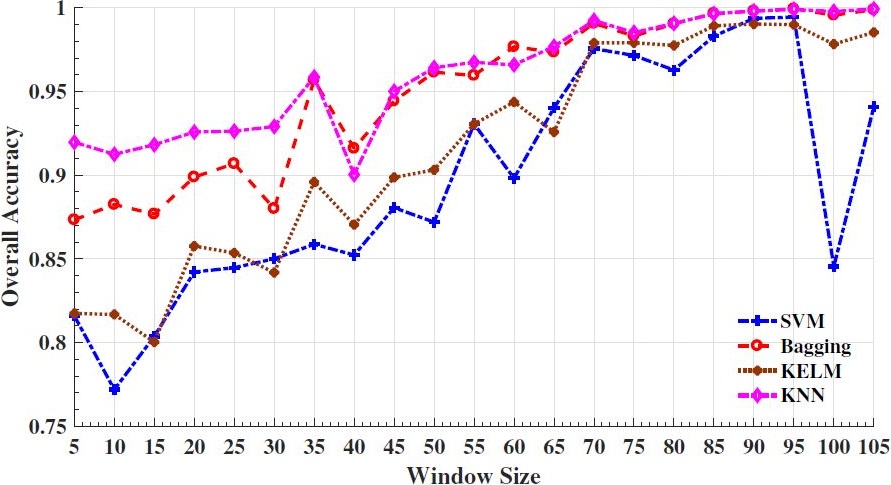}
    \caption{Cumulative Overall Accuracy across all the users for all classifiers.}
    \label{Fig.1A}
\end{figure}

\begin{figure}[!ht]
    \centering
    \includegraphics[scale=0.45]{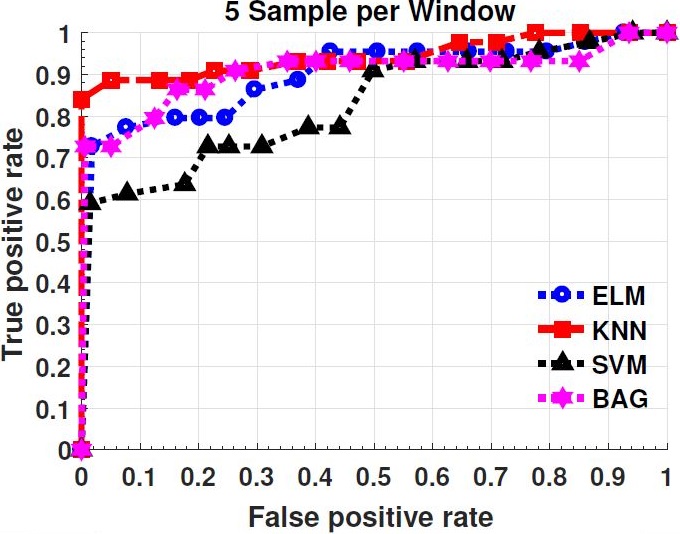}
    \includegraphics[scale=0.45]{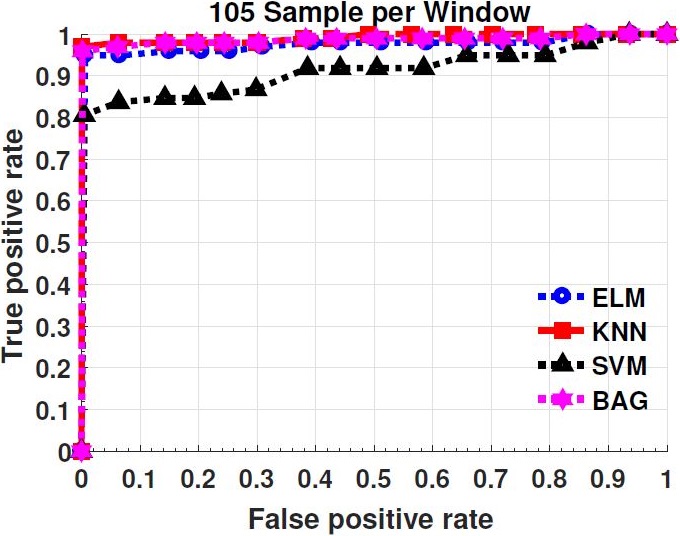}
    \caption{ROC for two different sample sizes, i.e., $5$ and $105$ samples per window. Here are the values of AuC's for each classifier: ELM = $0.8175$ ($5$ samples) and $0.9853$ ($105$ samples). Bagging = $0.8734$ ($5$ samples) and $0.9988$ ($105$ samples). SVM = $0.8161$ ($5$ samples) and $0.9408$ ($105$ samples). KNN = $0.9197$ ($5$ samples) and $0.9993$ ($105$ samples).}
    \label{Fig.2}
\end{figure}

In-terms of machine learning, the extensive model performance measurement is required. Moreover, when it comes to the classification, overall accuracy (as shown in Figure \ref{Fig.1A}) and the area under the curve, such as the receiver Operating Characteristics Curve (ROC) is an essential evaluation metric at various threshold settings. ROC is a probability curve that measures the degree of separability among classes. ROC curve is plotted with TPR against the FPR values obtained through the classification method. To validate the statistical significance of the proposed pipeline, the ROC has been drawn for two different sample sizes i.e., $5$ and $105$ samples per window. The results presented in Figure \ref{Fig.2} uphold the effectiveness of the proposed pipeline for real-time applicability with a $99\%$ confidence interval of legitimate user identification by using a pairwise T-test between a group of individuals. Looking at Figure \ref{Fig.2}, significant statistical results are seen; showing that all the classifiers outperform. Finally, we present the computational cost in terms of the time of enlisted experiments. Figure \ref{Fig.3} shows the computational time taken by the feature extraction and feature selection process for average users with a different number of samples per window.

As shown in Figure \ref{Fig.3}, the processing time is increased gradually at the start and then increased exponentially as the sample size increases. Therefore, to cope with the high computational time may become an important issue for the legitimate user identification system for larger sample sizes. There are many ways to overcome such high computational time, however, for this one needs to work on fewer features, i.e., either gait-based features or time and frequency domain feature processed through one of the feature selection methods but this may bring incompetency for statistical significance.

\begin{figure}[!ht]
\centering
	\begin{tikzpicture}
	\begin{axis}[height = 7cm, width = 8.0cm, grid = major,
    	title = Without legend box,
    	legend pos = north west, 
    	legend columns= 1,
        legend style = {draw = none},
	title={\textbf{Feature Extraction}},
	xlabel = \textbf{Samples per window},
	ylabel = \textbf{Time in Seconds},
	xmin = 0, xmax = 105, ymin = 0, ymax = 2300]
	\addplot[smooth, mark=*, blue] plot coordinates {
		(5,64.4331474) 
		(10,70.6142884)
		(15,75.3961687)
		(20,72.7416863)
		(25,79.5806419)
		(30,81.1719773)
		(35,86.6871829)
		(40,94.1907559) 
		(45,100.2960267)
		(50,109.8145144)
		(55,120.2552989)
		(60,136.3886357)
		(65, 148.7289928)
		(70,171.1221453)
		(75,198.2470186) 
		(80,233.9397287)
		(85,290.9363123)
		(90,388.8811129)
		(95,576.099192)
		(100,1136.273659)
		(105,2253.003516)
	};
	\end{axis}
	\end{tikzpicture} \ \ \ \ \
	\begin{tikzpicture}
	\begin{axis}[height = 7cm, width = 8.0cm, grid = major,
    	title = Without legend box,
    	legend pos = north west, 
    	legend columns= 1,
        legend style = {draw = none},
	title={\textbf{Feature Selection}},
	xlabel = \textbf{Samples per window},
	ylabel = \textbf{Time in Seconds},
	xmin = 0, xmax = 105, ymin = 0, ymax = 78200]
	\addplot[smooth, mark=*, blue]
	plot coordinates {
	    (5,30.2299496) 
		(10,30.7531847)
		(15,30.9892553)
		(20,33.7023063)
		(25,37.2552055)
		(30,40.821339)
		(35,46.8236344)
		(40,52.5360723) 
		(45,61.6753732)
		(50,72.6642161)
		(55,87.6603085)
		(60,108.3968503)
		(65,132.7678084)
		(70,177.5094689)
		(75,240.8478647) 
		(80,396.3056303)
		(85,714.8610704)
		(90,1327.07413)
		(95,3039.704188)
		(100,14792.19426)
		(105,78100.55629)
	};    
	\end{axis}
	\end{tikzpicture}
	\caption{\textbf{Processing Time} with different number of samples per window selected $([5:5:105])$ in each round for \textbf{Legitimate User Identification}. All the experiments are carried out on notebook using MATLAB (2017a) on Intel Core (TM) i5 CPU 2.40~GHz, with 8 GB RAM.}
	\label{Fig.3}
\end{figure}

\arrayrulecolor{black} 
\begin{table*}[!hbt]
    \caption{Cross subjects performance comparison with State-of-the-art works.}
    \centering
    \begin{tabular}{c|c|p{5cm}|p{5cm}|c}
        \textbf{Work} &  \textbf{Users} & \textbf{Features} & \textbf{Classifier} & \textbf{Accuracy} \\ \hline
        \cite{Ahmad2016} & 2 & Time Domain & SVM and KNN & 90\% \\ \hline
        \cite{Ahmad19} & 6 & Time and Frequency & Decision Tree, KNN, SVM & 98\%  \\ \hline 
        \cite{Ahmad2019} & 20 & Time and Frequency & Extreme Learning Machine & 97\% \\ \hline 
        \cite{Davidson16} &  10 & Gait, Time and Frequency & KNN and Random Forest & 90\% \\ \hline
        \cite{Kobayashi11} & 58 & Cross-correlations of Fourier transform & Nearest means in Fisher discriminant space and majority voting & 50\% \\ \hline  
        \cite{Thang12} & 11 & Time and Frequency & Gait templates, DTW, SVM & 93\% \\ \hline 
        \cite{Wolff13} & 36 & Variance in acceleration and orientation across $x, y, z$ & Gaussian distribution model & 83\% \\ \hline
        \cite{Sprager09} & 6 & Acceleration & SVM & 93\% \\ \hline
        \cite{Kwapisz10} & 36 & Time domain & J48 and ANN & 93\% \\ \hline
        \cite{Lin14} & 10 & Spectral energy diagrams of pitch, roll, acceleration $x, y, z$ & $\alpha \beta$ filtering, Empirical Mode Decomposition, Fourier Transform, Linear Discriminant Analysis & 90\% \\ \hline 
        \cite{Hughes16} & 2  &  \multicolumn{2}{c|}{Genetic Programming} & 90\%\\ \hline  
        \cite{Derawi13} & 10 & Magnitude of the acceleration Weighted moving average filter, cycle detection, Manhattan distance & SVM & 88\% \\ \hline
        \cite{Pan09} & 30 & Extrema in acceleration  & Difference-of-Gaussian and KNN & 96\% \\ \hline
        \textbf{Proposed} & 16 & Time, Frequency and Gait & Extreme Learning Machine  & $98\%$ \\ \hline 
        \textbf{Proposed} & 16 & Time, Frequency and Gait & Support Vector Machine & $94\%$ \\ \hline 
        \textbf{Proposed} & 16 & Time, Frequency and Gait & K Nearest Neighbour  & $99\%$\\ \hline 
        \textbf{Proposed} & 16 & Time, Frequency and Gait & BAG & $99\%$ \\ \hline 
    \end{tabular}
    \label{Tab.20}
\end{table*}

\begin{table*}[!ht]
\captionsetup{justification=centering}
\caption{Average Accuracy, Confidence Intervals and Time taken for Legitimate User Identification for \(50\) Sample Per Window With Different Feature Selection Methods and Different Number of Features for two different sensors data.} 
\centering 
\resizebox{\textwidth}{!}{\begin{tabular}{c|c|cccccccccccccccc}
\hline

\multirow{4}{*}{\bf Features}& \multirow{4}{*}{\bf Metric}  & \multicolumn{4}{|c}{\bf Back Left Pocket} & \multicolumn{4}{|c}{\bf Back Right Pocket} & \multicolumn{4}{|c}{\bf Front Left Pocket}& \multicolumn{4}{|c}{\bf Back Right Pocket} \\ [1.0ex]

& & \multicolumn{2}{|c}{\bf ACC}&\multicolumn{2}{|c}{\bf LACC}&\multicolumn{2}{|c}{\bf ACC}& \multicolumn{2}{|c}{\bf LACC}&\multicolumn{2}{|c}{\bf ACC}&\multicolumn{2}{|c}{\bf LACC}&\multicolumn{2}{|c}{\bf ACC}&\multicolumn{2}{|c}{\bf LACC} \\ [1.0ex]

&& \multicolumn{1}{|c}{\bf PCA}&\multicolumn{1}{|c}{\bf ESMP}&\multicolumn{1}{|c}{\bf PCA}&\multicolumn{1}{|c}{\bf ESMP}&\multicolumn{1}{|c}{\bf PCA}&\multicolumn{1}{|c}{\bf ESMP}&\multicolumn{1}{|c}{\bf PCA}&\multicolumn{1}{|c}{\bf ESMP}&\multicolumn{1}{|c}{\bf PCA}& \multicolumn{1}{|c}{\bf ESMP}&\multicolumn{1}{|c}{\bf PCA}&\multicolumn{1}{|c}{\bf ESMP}&\multicolumn{1}{|c}{\bf PCA}&\multicolumn{1}{|c}{\bf ESMP}&\multicolumn{1}{|c}{\bf PCA}& \multicolumn{1}{|c}{\bf ESMP}
\\ [2.5ex]
\hline

& \bf Accuracy &55\(\pm\)5.1&50\(\pm\)3.2&50\(\pm\)4.5&53\(\pm\)5.7&52\(\pm\)4.9&53\(\pm\)3.8&51\(\pm\)3.6&50\(\pm\)2.7&54\(\pm\)3.6&55\(\pm\)2.9&57\(\pm\)2.9&54\(\pm\)3.9&54\(\pm\)4.6&63\(\pm\)2.9&54\(\pm\)3.8&52\(\pm\)4.5 \\[-1ex]
\raisebox{1.5ex}{\bf 5}  \raisebox{1.5ex} & \bf Time
&0.140&0.138&0.209&0.243&0.265&0.261&0.553&0.527&0.175&0.173&	0.268&0.270&0.164&	0.133&0.201&0.202\\ [1.0ex]
\hline

& \bf Accuracy &74\(\pm\)6.3&85\(\pm\)3.3&67\(\pm\)4.8&85\(\pm\)3.8&63\(\pm\)4.5	&77\(\pm\)2.3	&63\(\pm\)3.4	&84\(\pm\)3.9	&66\(\pm\)2.9	&90\(\pm\)4.6	&72\(\pm\)4.1	&81\(\pm\)3.9	&69\(\pm\)3.8	&74\(\pm\)5.9	&64\(\pm\)4.7	&77\(\pm\)4.8\\[-1ex]
\raisebox{1.5ex}{\bf 10}  \raisebox{1.5ex} & \bf Time
&0.140	&0.142	&0.209	&0.214	&0.255	&0.252	&0.535	&0.523	&0.177	&0.178	&0.270	&0.271	&0.131	&0.131	&0.194	&0.188\\[1ex]
\hline

& \bf Accuracy &72\(\pm\)6.2&	98\(\pm\)1.1&	75\(\pm\)4.6&	89\(\pm\)4.1&	76\(\pm\)5.5&	99\(\pm\)0.9&	78\(\pm\)3.9&	96\(\pm\)2.0&	70\(\pm\)5.8&	95\(\pm\)3.4&	77\(\pm\)4.1&	95\(\pm\)2.1&	74\(\pm\)4.1&	98\(\pm\)1.3&	71\(\pm\)5.7&	98\(\pm\)1.0\\[-1ex]
\raisebox{1.5ex}{\bf 15}  \raisebox{1.5ex} & \bf Time
&0.142	&0.142	&0.211	&0.238	&0.255	&0.254	&0.536	&0.517	&0.177	&0.178	&0.269	&0.273	&0.135	&0.132	&0.197	&0.203\\[1ex]
\hline

& \bf Accuracy &73\(\pm\)6.1&	97\(\pm\)1.4&	73\(\pm\)4.9&	97\(\pm\)2.0&	76\(\pm\)4.6&	99\(\pm\)0.4&	82\(\pm\)2.5&	98\(\pm\)1.6&	76\(\pm\)3.6&	99\(\pm\)0.7&	76\(\pm\)6.7&	98\(\pm\)2.6&	71\(\pm\)7.1&	95\(\pm\)2.4&	73\(\pm\)5.8&	94\(\pm\)4.2\\[-1ex]
\raisebox{1.5ex}{\bf 20}  \raisebox{1.5ex} & \bf Time
&0.142	&0.142	&0.216	&0.219	&0.259	&0.256	&0.538	&0.529	&0.178	&0.189	&0.269	&0.281	&0.132	&0.132	&0.194	&0.212\\[1ex]
\hline 

& \bf Accuracy &78\(\pm\)3.8&	99\(\pm\)0.5&	74\(\pm\)4.9&	98\(\pm\)2.7&	72\(\pm\)4.4&	96\(\pm\)3.4&	80\(\pm\)4.6&	98\(\pm\)1.6&	75\(\pm\)6.7&	98\(\pm\)1.2&	76\(\pm\)4.9&	94\(\pm\)2.3&	73\(\pm\)6.7&	96\(\pm\)3.9&	73\(\pm\)8.3&	90\(\pm\)4.9\\[-1ex]
\raisebox{1.5ex}{\bf 25}  \raisebox{1.5ex} & \bf Time
&0.148	&0.140	&0.221	&0.219	&0.258	&0.257	&0.532	&0.528	&0.180	&0.179	&0.254	&0.270	&0.134	&0.133	&0.200	&0.203 \\[1ex]
\hline 

& \bf Accuracy & \bf 78\(\pm\)3.8&	\bf 97\(\pm\)0.4&	\bf 76\(\pm\)4.3&	\bf 98\(\pm\)1.8&	\bf 73\(\pm\)6.9&	\bf 98\(\pm\)1.5&	\bf 76\(\pm\)3.6&	\bf 98\(\pm\)1.4&	\bf 76\(\pm\)6.7&	\bf 94\(\pm\)3.2&	\bf 76\(\pm\)6.4&	\bf 99\(\pm\)0.3&	\bf 74\(\pm\)3.7&	\bf 99\(\pm\)0.5&	\bf 68\(\pm\)7.8&	\bf 91\(\pm\)4.1\\[-1ex]
\raisebox{1.5ex}{\bf 30}  \raisebox{1.5ex} & \bf Time
& \bf 0.148	& \bf 0.143	& \bf 0.216	& \bf 0.217	& \bf 0.260	& \bf0.258	& \bf 0.539	& \bf 0.532	& \bf 0.179	& \bf 0.179	& \bf 0.268	& \bf 0.284	& \bf 0.132	& \bf 0.149	& \bf 0.197	& \bf 0.181 \\[1ex]
\hline 

& \bf Accuracy &77\(\pm\)5.4&	99\(\pm\)0.5&	71\(\pm\)6.2&	94\(\pm\)4.7&	72\(\pm\)4.5&	99\(\pm\)0.3&	74\(\pm\)5.9&	87\(\pm\)6.2&	75\(\pm\)4.5&	97\(\pm\)1.7&	73\(\pm\)5.5&	97\(\pm\)3.6&	76\(\pm\)7.2&	99\(\pm\)0.8&	69\(\pm\)7.9&		54\(\pm\)4.9\\[-1ex]
\raisebox{1.5ex}{\bf 35}  \raisebox{1.5ex} & \bf Time
&0.143	&0.143	&0.216	&0.208	&0.258	&0.259	&0.536	&0.548	&0.180&	0.179&	0.267	&0.275	&0.134&	0.133	&0.194	&0.201 \\[1ex]
\hline 

& \bf Accuracy & \bf 74\(\pm\) \bf 5.9&	 \bf 98\(\pm\) \bf 0.6&	 \bf 70\(\pm\) \bf 6.4&	 \bf 99\(\pm\) \bf 0.4&	 \bf 71\(\pm\) \bf 4.2&	 \bf 98\(\pm\) \bf 0.8&	 \bf 72\(\pm\) \bf 5.9&	 \bf 99\(\pm\) \bf 0.4&	 \bf 73\(\pm\) \bf 4.6&	 \bf 50\(\pm\) \bf 2.8&	 \bf 71\(\pm\) \bf 4.9&	 \bf 75\(\pm\) \bf 5.9&	 \bf 69\(\pm\) \bf 8.2&	 \bf 98\(\pm\) \bf 1.1&	 \bf 69\(\pm\) \bf 6.3&	 \bf 99\(\pm\) \bf 0.4\\[-1ex]
\raisebox{1.5ex}{\bf 40}  \raisebox{1.5ex} & \bf Time
& \bf 0.144&	 \bf 0.142&	 \bf 0.216&	 \bf 0.212&	 \bf 0.262&	 \bf 0.259&	 \bf 0.537&	 \bf 0.532&	 \bf 0.180&	 \bf 0.172&	 \bf 0.276&	 \bf 0.278&	 \bf 0.133&	 \bf 0.148&	 \bf 0.196&	 \bf 0.196 \\[1ex]
\hline 
\end{tabular}}
\label{Tab.12A}
\end{table*}

From experimental results, one can conclude that the ESMP helps to boost legitimate user identification performance. Based on the results listed in Figure \ref{Fig.1A} and Tables \ref{Tab.8}-\ref{Tab.9}, we observe that the ESMP together with almost all classifiers works better and accurately than the several state-of-the-art legitimate user identification methods. The experiments show the process of analyzing the behavior of a different number of samples per window taken by the user for legitimate user identification (i.e., $05: 05: 105$). Tables \ref{Tab.12}-\ref{Tab.15} present the accuracy of our proposed pipeline for an individual user being identified correctly with the different number of samples per window after the fusion of four different built-in sensors data. From the results, we found that the proposed pipeline produces acceptable results with $30-50$ samples per window. Figure \ref{Fig.1A} enlists the results of overall accuracy across an all user with a $99\%$ confidence interval.

\section{Comparison against State-of-the-art Solutions}
\label{sec:5}

Hereby we present and compare some of the critical works from the literature \cite{Ahmad2016, Ahmad19, Ahmad2019, Davidson16, Kobayashi11, Thang12, Wolff13, Sprager09, Kwapisz10, Lin14, Hughes16, Derawi13, Pan09} that can be categorized in two groups such as implicit and multi-modality biometrics. All these methods have some limitations i.e., required some additional information and source or may require user interaction. Therefore, to some extent, all these methods are innovative but require some external legitimate user identification process.

The works compared in Table \ref{Tab.20} require user interaction in-terms to perform a predefined activity or user data has been gathered in controlled environments which are not a real representation of frequent user interactions. Therefore, the proposed method could be considered as an exciting alternative for continuous and explicit legitimate user identification or impostor identification in a semi-controlled environment. Hence the proposed method overcomes the limitations of the smartphone in power consumption and user interaction. We have worked on combining different feature extraction (time, frequency, and gait) and selection (ESMP) techniques and to concatenating the selected features to deliver a reliable legitimate user identification model using built-in sensors data in a semi-controlled environment. 

As we earlier discussed, this study focuses on the idea of identifying a smartphone user by applying different walking patterns. Furthermore, it is assumed that the phone is freely placed without any particular orientation inside any of the user's pants pockets (\textit{front left, front right, back left, back right}). Thus to answer the questions "Does the ESMP, a non-linear unsupervised feature selection method improve the identification accuracy more than the other existing and well-studied unsupervised feature selection methods such as Principal Component Analysis (PCA)?" and "Does the data variation affect the performance of the LUI process?". Here we enlist the comparative analysis of these two different feature selection methods with a different number of features and different locations of a smartphone. The experimental results are shown in Table \ref{Tab.12A}. One can conclude that ESMP significantly performed better than PCA. 
\section{Conclusion}
\label{sec:6}

Smartphones are becoming increasingly popular that have forced the community to study the security implications of these devices. This work suggested that gait-based legitimate user identification is possible in an uncontrolled environment with some limitations. The proposed pipeline has some attractive features to its applicability such as smaller confidence intervals that imply more reliability in training. Furthermore, holding a permanent structure is a useful feature for hardware constraints such as transforming the trained model into a chip which can further increase device security by not allowing software-based attacks but only hardware manipulations. These hardware operations would require access to the smartphone hence making such attacks subject to the device defense.

The proposed pipeline achieves a $100\%$ true positive and $0\%$ false-negative rate for almost all classifiers. However, to further validate the claims, it may be useful to check the sensor quality while changing the smartphone as well as with different operating systems. 

The key advantage of our study is that the samples for each user are collected on different days with different jeans, locations, and orientation which significantly helps to understand the characteristic behavior of users which is an essential component for any legitimate user identification system. However, one of the main limitations of gait-based legitimate user identification is that the signals inconsistency e.g., signal reliability, degrades significantly between days due to many factors such as a change in habits, mood, workload, etc. which we will address in our future studies.

\bibliographystyle{IEEEtran}
\bibliography{samp}
\end{document}